\documentclass[usegraphicx,usenatbib]{mn2e}

\def\ni{\noindent}

\def\rysunek#1#2#3{
\begin{figure}
\includegraphics[width=88mm]{#1}
\caption{#2}
\label{#3}
\end{figure}
}

\def\poczwrys#1#2#3#4#5#6{
\begin{figure}
\vbox{
\includegraphics[width=40mm]{#1}\hskip 4mm\includegraphics[width=40mm]{#2}
\hfill\break
\includegraphics[width=40mm]{#3}\hskip 4mm\includegraphics[width=40mm]{#4}
}
\caption{#5}
\label{#6}
\end{figure}
}

\def\poszrys#1#2#3#4#5#6#7#8{
\begin{figure}
\vbox{
\includegraphics[width=40mm]{#1}\hskip 4mm\includegraphics[width=40mm]{#2}
\hfill\break
\includegraphics[width=40mm]{#3}\hskip 4mm\includegraphics[width=40mm]{#4}
\hfill\break
\includegraphics[width=40mm]{#5}\hskip 4mm\includegraphics[width=40mm]{#6}
}
\caption{#7}
\label{#8}
\end{figure}
}

\title{The First Compact Objects in the $\Lambda$-dominated Universe}
\author[S.~Stachniewicz, M.~Kutschera]
{S.~Stachniewicz$^1$, M.~Kutschera$^{1,2}$\\
$^1$Astrophysics Division, H.Niewodnicza\'nski Institute of Nuclear Physics,
ul. Radzikowskiego 152,\\
31-342 Krak\'ow, Poland\\
$^2$Institute of Physics, Jagiellonian University, ul. Reymonta 4,\\
30-059 Krak\'ow, Poland
}
\date{}

\begin{document}
\maketitle

\begin{abstract}

We calculate the evolution of a low-mass ($M \le 10^5 M_{\odot}$) spherically
symmetric density perturbation in the $\Omega_b h^2=0.02$, $\Omega_M =0.35$,
$\Omega_{\Lambda}=0.65$, $h=0.72$ Universe. The results are compared with
the ones that assume no cosmological constant and the flat, dark matter
dominated Universe. We include thermal processes and non-equilibrium chemical
evolution of the collapsing gas. We find that
direct formation of bound objects with such masses by $z=8$ is
unlikely
so in fact they may form only through fragmentation of greater objects. This
is in stark contrast to the $\Omega=1$ pure CDM cosmology, where low-mass
objects form abundantly at redshits $z>10$.
\end{abstract}


\section{Introduction}

Present observations of the CMB anisotropies by the BOOMERANG \citep{Net02},
MA\-XI\-MA-1 \citep{Lee01}, CBI \citep{Pad01} and DASI \citep{Pry02}
experiments combined with the data from the Supernova project
(\citet{Rie98}; \citet{Per99}) suggest that although
the geometry of the Universe is flat, its matter content is dominated by the
vacuum energy or so-called quintessence. The background
cosmology has a very strong influence on the structure formation. In this
paper we address the formation of first bound objects in the flat $\Lambda$CDM
cosmology.

  The structure
formation on both large and small scales
has been studied very carefully
in the dark matter dominated cosmology, in particular in the flat pure CDM
model. This is not so for
the vacuum-energy-dominated models, which are of prime relevance according to
the above observations.
Perhaps one of the most interesting objects to study are the first
bound objects in the Universe -- the first quasars and the Population III
stars that were the first sources of photons after the light of CMB faded
away, and ended
the `Dark Ages' of the Universe. Some of these calculations were described
in a review by \citet{Bar01}. The recent identification of quasars at $z\sim6$
in the Sloan Digital Sky Survey \citep{Fan01} strongly motivates such an
investigation.

Our code is similar to the one by \citet{Tho95} and by \citet*{Hai96}.
We consider the evolution of
a single spherically symmetric density perturbation  in the early Universe
starting soon after recombination until it finally forms a bound stationary
object. Our aim here is to extend their analysis, which was restricted to the
$\Omega=1$ CDM cosmology, to the $\Lambda$-dominated
Universe and compare the results with  those for the flat, dark matter
dominated
Universe but with
more up-to-date cosmological parameters ($h=0.72$, $\Omega_b h^2=0.02$).

We start tracing the initial expansion of the perturbation at high redshift
when its density contrast is still in the linear regime. Then we follow
decoupling of the
perturbation  from the Hubble flow and its subsequent collapse and formation
of a virialized cloud. We include the gas dynamics and various cooling and
heating
processes operating in the expanding and collapsing cloud. The chemical
evolution of the collapsing primordial gas cloud is also accounted for.

After the initial collapse, a virialized gas cloud is formed. The kinetic
energy of the infalling gas is dissipated through shocks and the cloud becomes
pressure-supported. We study the further evolution of the cloud which is
determined by its ability to cool sufficiently fast. The most important
cooling mechanism for low-mass clouds is the radiation of excited H$_2$
molecules. The presence of a small amount of the molecular hydrogen H$_2$
is crucial for triggering the final collapse of such clouds which could form
the first luminous object in the Universe.

\section{Nonlinear evolution of a spherically symmetric density perturbation}

A major approximation we use is the assumption of
spherical symmetry. This assumption is best justified in the case
of first objects formed in the Universe which are supposed to originate from
the rare highest fluctuations in the primordial density field
\citep{Bar86}. It allows us to focus mainly on gas dynamical processes which
are expected to control
the collapse of low-mass clouds which are supposed to form the first luminous
objects.

 The spectrum
of density fluctuations has more power on small scales, hence the first
nonlinear structures are expected to occur on relatively small scales
\citep{MiC99}.
As we are interested in small scales, much lower than the horizon, we use
Newtonian
gravity and treat the expansion of the Universe as a hydrodynamical flow.
To describe  the evolution of a spherically symmetric density
perturbation in the nonlinear regime we use Lagrangian coordinates. We
divide both baryonic and dark matter into concentric shells. The dynamics of
primordial matter is given by the equations below.

 The continuity equation for baryonic matter reads

\begin{equation} {dM_b \over dr_b}=4\pi r_b^2 \varrho_b, \label{ciaglosc}
\end{equation}

\ni where $r_b$ is the radius of a sphere of mass $M_b$. The radial velocity
of
the surface of this sphere is

\begin{equation} {dr_b \over dt} = v_b , \label{promien}\end{equation}

\ni and the acceleration satisfies the dynamical equation

\begin{equation}
{dv_b \over dt}= -4\pi r_b^2 {dp \over dM_b}-{GM(r_b) \over r_b^2} ,
\label{predkosc}
\end{equation}

\ni where $M(r_b)=M_b(r_b)+M_{dm}(r_b)+2(M_{rad}(r_b)-M_{\Lambda}(r_b)$) is
the
total mass within radius $r_b$ including the effects of radiation and
vacuum energy components of the stress-energy tensor. Two last contributions
result from the CMB and cosmological
constant components of the energy density of the Universe,

\begin{equation}
M_{rad}(r_b) = {H_0^2 \over 2 G} \Omega_{rad} (z+1)^4 r_b^3
\end{equation}

\ni and

\begin{equation}
M_{\Lambda}(r_b) = {H_0^2 \over 2 G} \Omega_{\Lambda} r_b^3,
\end{equation}

\ni where $\Omega_{rad}=8\pi G a {T_{CMB}}^4/(3 H_0^2 c^2)$ is the present
contribution of the CMB to the total energy density of the Universe in terms
of the critical density. Although, as we have shown in \citep*{Sta01}, at least
the radiation component is not negligible, both dark energy and radiation
terms are not very important.

The energy conservation condition for baryonic matter reads

\begin{equation}
{du \over dt}={p \over \varrho_b^2} {d \varrho_b \over dt} + {\Lambda_{cool}
\over \varrho_b} ,
\label{energia}
\end{equation}

\ni where $u$ is the internal energy per unit mass, $p$ is the pressure and
$\varrho_b$ is the baryon density. The last term in the eq.(\ref{energia})
describes cooling/heating of the gas, with $\Lambda_{cool}$ being the energy
emission (absorption) rate per unit volume.

We use the equation of state of the ideal gas

\begin{equation} p= (\gamma -1) \varrho_b u , \end{equation}

\ni where $\gamma = 5/3$, as the primordial baryonic matter after
recombination is assumed to be composed
of monoatomic hydrogen and helium with a small (but very important) admixture
of molecular
hydrogen H$_2$.

Dynamics of dark matter is simpler, as we assume it to be collisionless.
The continuity equation is

\begin{equation} {dM_{dm} \over dr_{dm}}=4\pi r_{dm}^2 \varrho_{dm} ,
\end{equation}

\ni where $r_{dm}$ is the radius of dark matter sphere of mass $M_{dm}$.
The radial velocity of
this sphere is

\begin{equation} {dr_{dm} \over dt} = v_{dm} , \end{equation}

\ni and the acceleration reads

\begin{equation} {dv_{dm} \over dt}= -{GM(r_{dm}) \over r_{dm}^2} .
\label{predkoscd}
\end{equation}

To solve the above equations we must specify the cooling/heating function
$\Lambda_{cool}$ and the the initial conditions.

\section{Initial conditions}

Let us begin with the initial conditions.
We start to follow the evolution of the perturbation soon after the
recombination at a sufficiently high
redshift that the perturbation is still in the linear regime. We choose the
initial redshift to be $z_i=500$ like in \citet{Hai96}.

We apply the initial density profiles in the form of a single spherical
Fourier mode used also by \citet{Hai96}

\begin{equation} \varrho_i(r)=\Omega_i \varrho_c (1+\delta_i {\sin kr \over
kr})
,
\label{psinus} \end{equation}

\ni where $i=b,dm$ and $\varrho_c$ is the critical density of the Universe,
$\varrho_c=3H^2/8\pi G$ with $H$ being the actual value of the Hubble
parameter.
The quantities $\delta_b$ and $\delta_{dm}$ measure, respectively, the baryon
and dark matter density enhancement with respect to the mean densities
${\bar \varrho_b}=\Omega_b\varrho_c$ and ${\bar \varrho_{dm}}=
\Omega_{dm}\varrho_c$.

For this profile there exist two distinguished values of the radius,
$R_0$ and $R_z$ which correspond, respectively, to the first zero and the
first minimum of the function $\sin (kr)/kr$. They are shown in Fig.
\ref{rozklsin}.

\rysunek{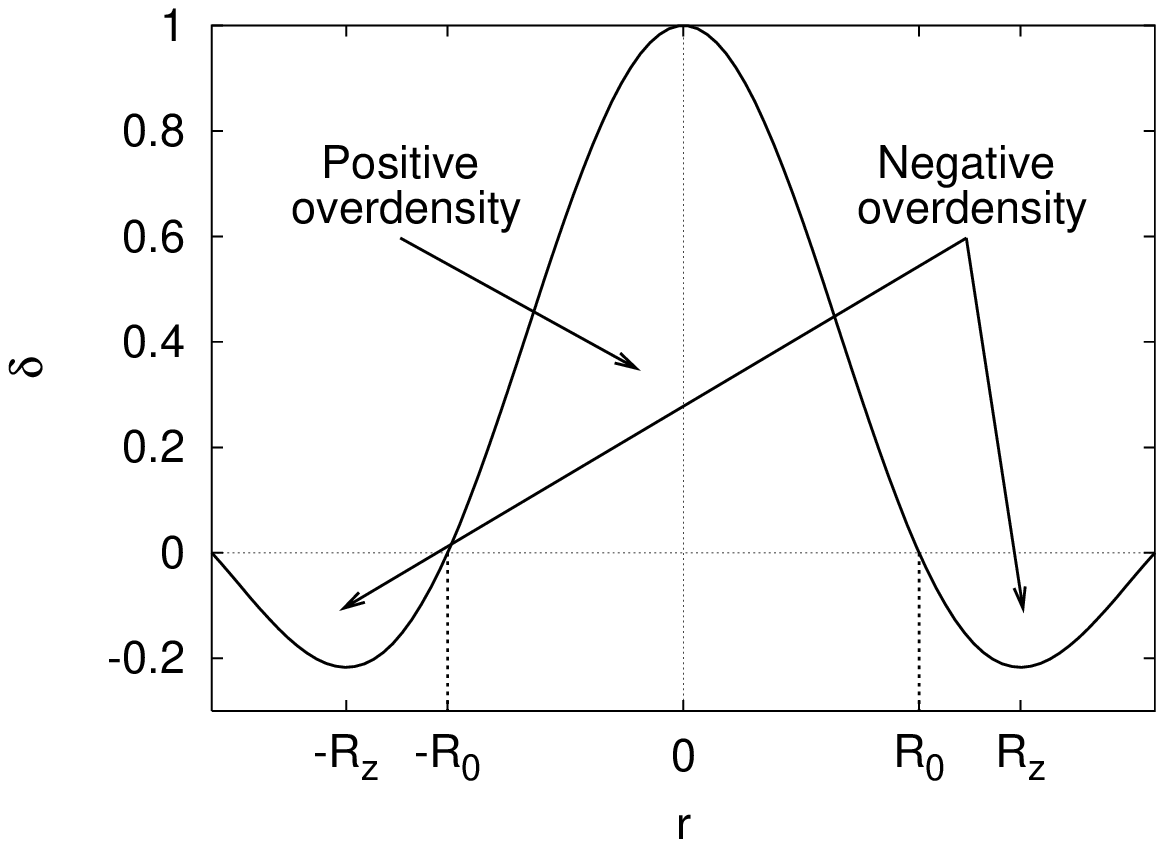}
{The `Sinusoidal' profile.}{rozklsin}

Inside the sphere of radius $R_0=\pi/k$ which contains mass
$M_0$, the local density contrast is positive. The
mass $M_0$ and the radius $R_0$ will be referred to as the cloud
mass and the cloud radius, respectively. The local
density contrast is negative for $R_z>r>R_0$, with the average
density contrast vanishing for the sphere of radius $R_z=4.49341/k$ with
the mass $M_z$. According to the gravitational instability
theory in the expanding Universe, the shell of radius $R_z$ will
expand together with the Hubble flow not suffering any additional deceleration.
This is why we regard this profile as very convenient in numerical simulations
because it eliminates numerical edge effects and mentioned shell simply follows
the Hubble expansion of the Universe.
It can thus be regarded as the boundary of the perturbation
and the mass $M_z$ will be referred to as the bound mass.

It is worth to note that for radii not greater than $3/4 R_0$ this profile
is very similar to the Gaussian profile

\begin{equation} \varrho_i(r)=\Omega_i \varrho_c \left[ 1+\delta_i \exp \left(
{-r^2 \over 2R_f^2} \right) \right] \end{equation}

\ni with $R_f=1/2 R_0$. Both profiles are compared in Fig. \ref{rozklady}.

\rysunek{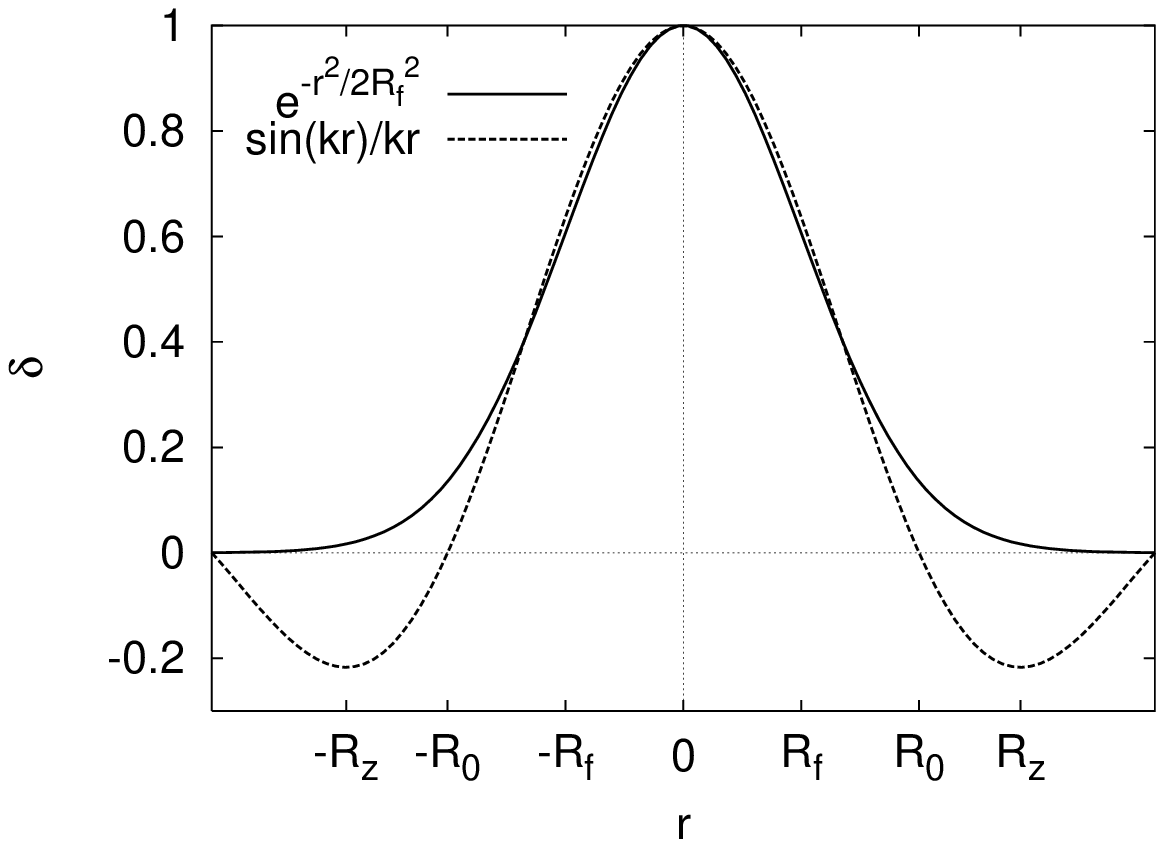}
{Comparison of the `Sinusoidal' profile and the Gaussian one.}{rozklady}

As the initial velocity we use the Hubble velocity for baryon matter,

\begin{equation}v_b(r)=Hr , \end{equation}

\ni whereas for dark matter the initial velocity is \citep{Hai96}

\begin{equation} v_{dm}(r)=Hr(1-{1 \over 3}<\delta_{dm}>_r) .\end{equation}

\ni The expression in brackets  indicates averaging over the sphere
of radius r.
\ni Slower expansion of dark matter at $z_i=500$ results from the fact that
the dark matter perturbations  start to grow  earlier  than do
the baryon matter ones.

Finally, the amplitudes of baryon and dark matter perturbations, $\delta_b$
and $\delta_{dm}$ should be specified. To do so one should use  the power
spectrum corresponding to the initial redshift $z_i=500$ that provides rms
density fluctuations at a given mass $M_0$. For Cold Dark Matter model we use
transfer functions calculated using the CMBFAST program by \citet{Sel96}.
The actual value of the rms density fluctuation is equal to

\begin{equation}
\sigma(R,z)=\sqrt{ \int\limits_0^{\infty} {dk \over k} P(k) \tilde{W}(k,R)^2 }
\end{equation}

\ni where $P(k)=4\pi d^2_{norm} 100 k^n 0.05^{1-n} k^3 t_f(k,z)^2$. The value
of normalization constant $d^2_{norm}$ may be also obtained from the CMBFAST
program.

The rms fluctuation $\sigma(R,z)$ depends on the shape of the window over
which the density is averaged. However, this dependence is quite weak.
In the table \ref{sigma} we list some values for $z=500$ obtained by applying
this formula to the transfer function and COBE-normalized normalization
obtained from CMBFAST to the Gaussian profile of filter radius $R_f=1/2 R_0$.
Instead of $R_f$, we use M$_0$ which is equal to the total baryonic mass
within a sphere of the radius R$_0$.

\begin{table}
\caption{RMS overdensities for $\Lambda$CDM and flat CDM.}
\begin{center}
\begin{tabular}{ccc}
\hline
M$_0$[M$_{\odot}$] & $\sigma_{\Lambda CDM}$ & $\sigma_{\rm CDM}$ \\
\hline
$5\times 10^2$ & 0.0539 & 0.1437 \\
\hline
$1\times 10^3$ & 0.0518 & 0.1369 \\
\hline
$2\times 10^3$ & 0.0496 & 0.1303 \\
\hline
$5\times 10^3$ & 0.0469 & 0.1219 \\
\hline
$1\times 10^4$ & 0.0449 & 0.1159 \\
\hline
$2\times 10^4$ & 0.0429 & 0.1099 \\
\hline
$5\times 10^4$ & 0.0404 & 0.1024 \\
\hline
$1\times 10^5$ & 0.0385 & 0.0968 \\
\hline
\end{tabular}
\end{center}
\label{sigma}
\end{table}

It is worth to note that these results are not much different from
the results obtained by applying the fit by \citet{Eis99}.
However, our results are systematically greater by about 23-24\% for the
$\Lambda$-dominated model and by about 7-10\% for the pure CDM model.
Probably it is due to the fact that the mentioned fit is less accurate for
larger redshifts (especially for $z>30$) and greater $\Omega_b/\Omega_M$
ratios.

As the baryon density perturbations start to grow only after the
recombination,
we have decided to set its initial value to $\delta_{b,i}=0.1\delta_{dm,i}$
\citep{Hai96}. Let us note that CMBFAST predicts $\delta_{b,i}$ to be
about $0.2\delta_{dm,i}$ in both $\Lambda$ and dark matter dominated models.
However, as we have shown in \citet{Sta01}, if $\delta_{b,i}/\delta_{dm,i}$
is 0.2 or less the results are pretty much insensitive to this ratio,
runs with its values equal to 0.2, 0.1 and 0.0 were almost indistinguishible.

\section{Baryonic matter components, chemical reactions and thermal effects}

To specify the cooling/heating function $\Lambda_{cool}$ in Eq.(\ref{energia})
one
should include all relevant thermal and chemical processes in
the primordial gas. There are many papers discussing the most
important contributions to the function $\Lambda_{cool}$ -- see e.g.
\citet*{Kat96}.
All formulas may be found in \citet{Sta01}.

The primordial gas consists of neutral atoms and molecules, ions and free
electrons. In this paper we have taken into account nine species:
H, H$^-$, H$^+$, He, He$^+$, He$^{++}$, H$_2$, H$_2^+$ and e$^-$.
Some authors (e.g. \citet{Gal98}) include also deuterium and
lithium.

The abundance of various species can, generally, change with
time as the chemical reactions between species occur and the
ionization and dissociation
photoprocesses take place in the hot gas. The chemical reactions
include such processes as e.g. the ionization of hydrogen and helium
by electrons, the recombination of ions with electrons,
the formation of negative hydrogen ions, the formation of H$_2$
molecules, etc. The full list of relevant chemical reactions is
given in \citet{Sta01}.  Photoprocesses include ionization of
neutral hydrogen H, helium He and He$^+$ by photons, and
dissociation of negative hydrogen ions H$^-$ and H$_2$ molecules
by photons.

Time
evolution of the number density of the component $n_i$ is
described by the kinetic equation:

\begin{equation}
{dn_i \over dt} = \sum_{l=1}^9 \sum_{m=1}^9 a_{lmi} k_{lm} n_l n_m +
\sum_{j=1}^9 b_{ji} \kappa_j n_j .
\label{chemia}
\end{equation}

\ni The first component on the right-hand side of this equation describes
the chemical reactions and the other one describes photoionization and
photodissociation processes. The coefficients
$k_{lm}$ are reaction rates, the quantities $\kappa_n$ are photoionization
or photodissociation
rates and $a_{lmi}$ and $b_{ji}$ are numbers equal to 0, $\pm 1$ or $\pm 2$
depending on the reaction.
All the reaction rates, photoionization and photodissociation rates
are given in \citet{Sta01}.

The cooling (heating) function $\Lambda_{cool}$ includes emissivities
(absorption rates) due to such
processes as the collisional ionization of H, He and He$^+$,
recombination to H, He and He$^+$, the collisional excitation of
H and He$^+$, Bremsstrahlung, the Compton cooling and the cooling by
deexcitation of H$_2$ molecules.
The formulae for the heating/cooling contributions of various processes
are given in \citet{Sta01}.

\section{Numerical code and initial conditions}

The dynamical equations (\ref{ciaglosc})-(\ref{predkoscd}) are solved
numerically. At each
timestep also the chemical composition of the gas is updated by
solving Eq.(\ref{chemia}) and the appropriate value of the cooling function
$\Lambda_{cool}$ is calculated.
We based our numerical code on the code described by
\citet{Tho95}, which is the standard, second-order accurate,
Lagrangian finite-difference scheme. Details of the code are described in
\citet{Sta01}.

We handle timesteps and central
boundary conditions in the  way proposed by Thoul and Weinberg \cite{Tho95}.
For dark matter, we treat the center as a hard sphere of some "small" radius
$r_c$. In order not to affect the results of calculations the value of
$r_c$ should be much less than any other characteristic radius in the problem
but it should not be too small because smaller $r_c$ means worse energy
conservation and longer computation time.  We have chosen $r_c$
equal to the initial radius of the most innermost dark matter shell.
For the baryonic component, if a shell falls below $r_c$ we assume that it
has `collapsed' and fix its radius, temperature and chemical composition.

In the collapse of baryon matter one encounters the formation of shocks.
In our code shocks are treated with the artificial viscosity technique
\citep{Ric67}.

It is worth to mention that set of equations \ref{chemia} is very stiff and
may require very small timesteps. To deal with this problem we follow
\citet{Hai96} and we used the STIFBS routine from \citet{Pre96}.

We have performed two sets of calculations, for the $\Lambda$-dominated
Universe ($\Omega_b h^2=0.02$, $\Omega_M =0.35$, $\Omega_{\Lambda}=0.65$)
and for the pure CDM model ($\Omega_b h^2=0.02$, $\Omega_M =1.0$), with
$h=0.72$ in both cases. We started our calculations at $z=500$.
The initial gas temperature
and chemical composition were obtained by running our own program that starts
at $z=10000$ with equilibrium values and ends at a required redshift. We have
compared its predictions to the results of \citet{Gal98} and
the differences were not greater than 10\%. One should note note that these
authors have
included more species (e.g. deuterium and lithium) and some of their reaction
rates were slightly different. The difference between the temperatures
of the baryonic gas and the CBR turned out to be lower than 1\% at $z=500$.

For the $\Lambda$-dominated model we have performed runs with cloud masses
M$_0$ from 2000 to 10$^5$ M$_{\odot}$, for the pure CDM model cloud masses
were between 500 and 5000 M$_{\odot}$. For each cloud mass we have chosen
a few values of the initial dark matter overdensity between 0.07 and 0.40,
depending on the mass.

We have divided the baryonic component to 200 concentric shells of equal
mass, the dark matter component was divided to 1000 equal mass shells.
It turned out to be enough -- in Fig. \ref{pordokl} we compare the calculated
amount of
virialized and collapsed mass for the same initial conditions and
three different runs: with 200 baryonic and 1000 dark matter shells,
200 baryonic and 2000 dark matter shells and, finally, with 400 baryonic
and 2000 dark matter shells. The results are  the same within our numerical
accuracy.

\rysunek{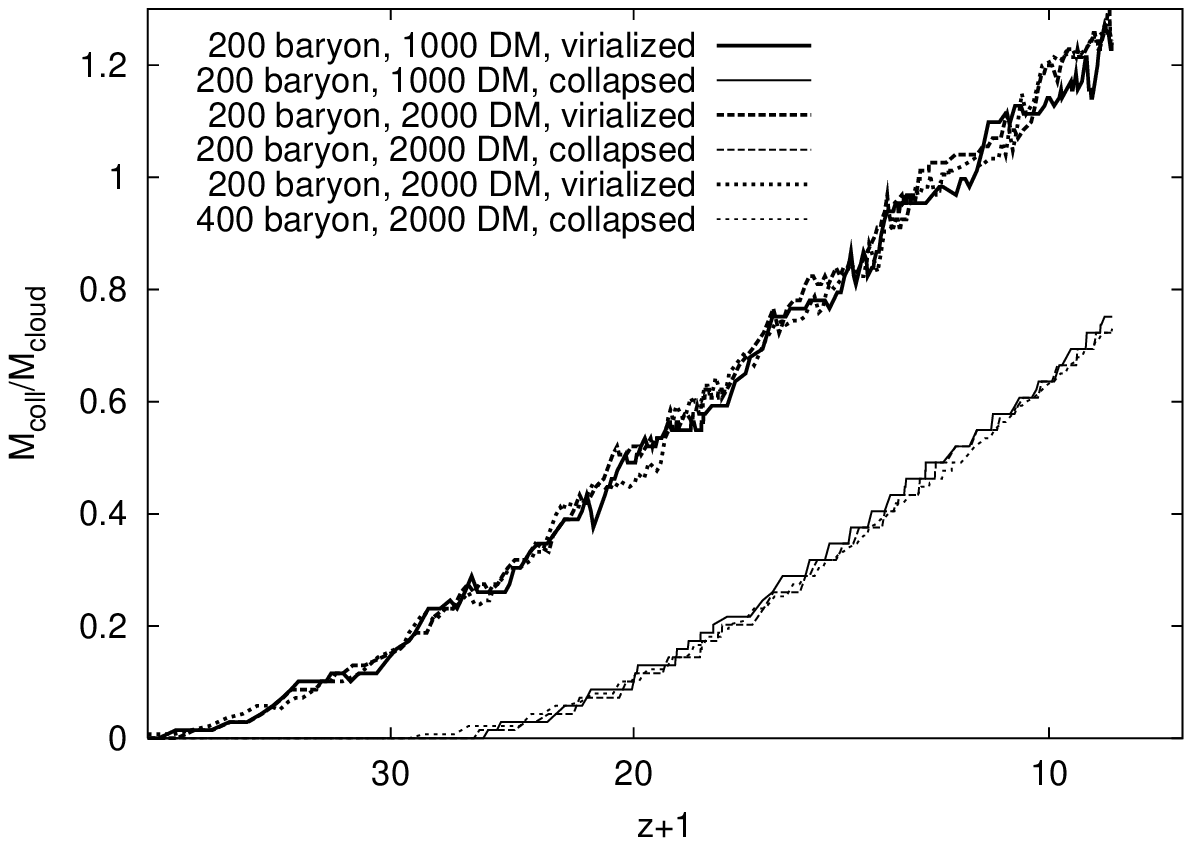}{Test of accuracy -- comparison of runs with various
number of baryonic and dark matter shells}{pordokl}

\section{Results}

We expect that there will be a significant difference between the low-mass
structure
formation in the $\Lambda$CDM and the Standard CDM models. The main reason
is that the rms fluctuations in the dark energy dominated model  are
about 2.5-2.7 times lower than in the Standard CDM model (table \ref{sigma}).
The other one is that in the Standard CDM the average
matter density is equal to the critical density so any overdensity would
eventually stop its expansion and collapse. In the $\Lambda$CDM the matter
density
is lower, so there is some threshold overdensity necessary for recollapse.
Any given overdensity would collapse slower (if at all) than in Standard CDM.
It means that in the $\Lambda$-dominated Universe there would be much less
luminous objects at large redshifts and the first ones would appear
much later as compared to the dark matter dominated Universe. In addition,
we may expect that the masses of the first collapsed objects would be
greater.

The behaviour of dark matter shells plays the crucial role in the cloud
collapse as there is much more dark matter than the baryonic
matter.
By assumption, dark matter is collisionless, so there
is no energy loss mechanism apart from possible gravitational energy exchange.
The evolution  of the dark matter shells in the absence of baryonic matter
is that their radii increase  to some maximal value, then the shells collapse
and after some oscillations due to bounces off the artificial hard sphere the
dark
matter cloud becomes stationary. In the presence of baryonic matter the
behaviour of dark matter shells is similar because the amount of baryonic
matter is much lower than the amount of dark matter. The difference is that
the average radius of a dark matter shell slowly decreases.

The behaviour of baryon matter shells is that after reaching maximum radii,
the shells collapse and the shock develops at some stage due to pressure
crowding of neighbouring shells \citep{Hai96}. The shock stops the collapse
and the baryon shells virialize. Then if the clouds have a sufficient
temperature and
density for efficient cooling, they can undergo the final collapse.

We assume that the shell has `virialized' if the mean density inside the
shell exceeds $18\pi^2$ times the actual mean baryon density of the Universe.
This value was calculated for the top-hat overdensities in the Einstein-de
Sitter
model but, however, \citet{Bry98} have shown that in the
$\Lambda$CDM models the dependence on $\Omega_M$ is not very strong.
Of course, this criterion depends on the redshift. For a virialized baryonic
shell its radius and density are roughly constant but the mean matter
density of the Universe decreases so the overdensity increases. This means
that overdensity is not a good criterion of the collapse -- we need to take
something related to the actual size of the object. We have decided to assume
that
a shell has `collapsed' if its radius falls below some `small' value,
in our case: the initial radius of the cloud $R_0$.

The results of our calculations are displayed in Figs. \ref{lambdacdm} --
\ref{txew}. Perhaps
the most interesting are plots that show the evolution of the amount of
the collapsed baryonic mass for various cloud masses and initial
overdensities. Figs. \ref{lambdacdm} a-d show the fraction of the cloud mass
that meets the collapse criterion at a given redshift
for the $\Lambda$-dominated Universe.
Similar plots for the dark matter dominated Universe are shown in Figs.
\ref{scdm} a-d.

\poczwrys{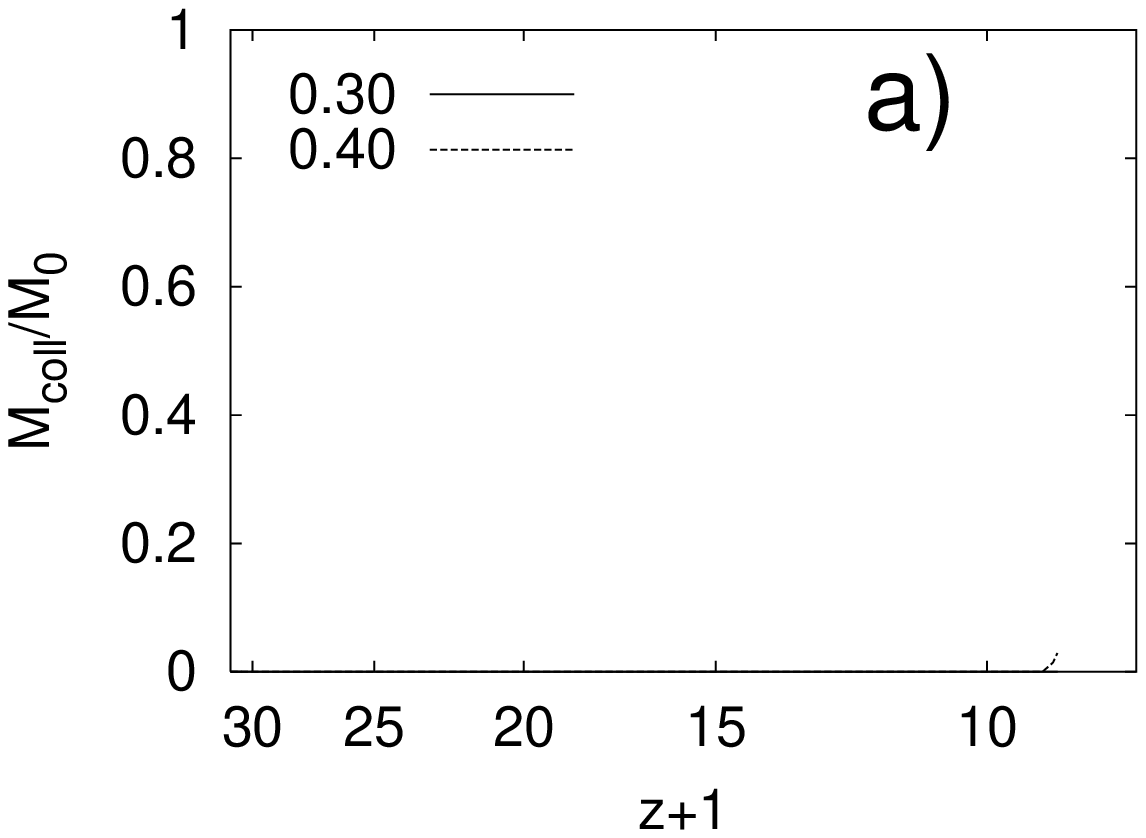}{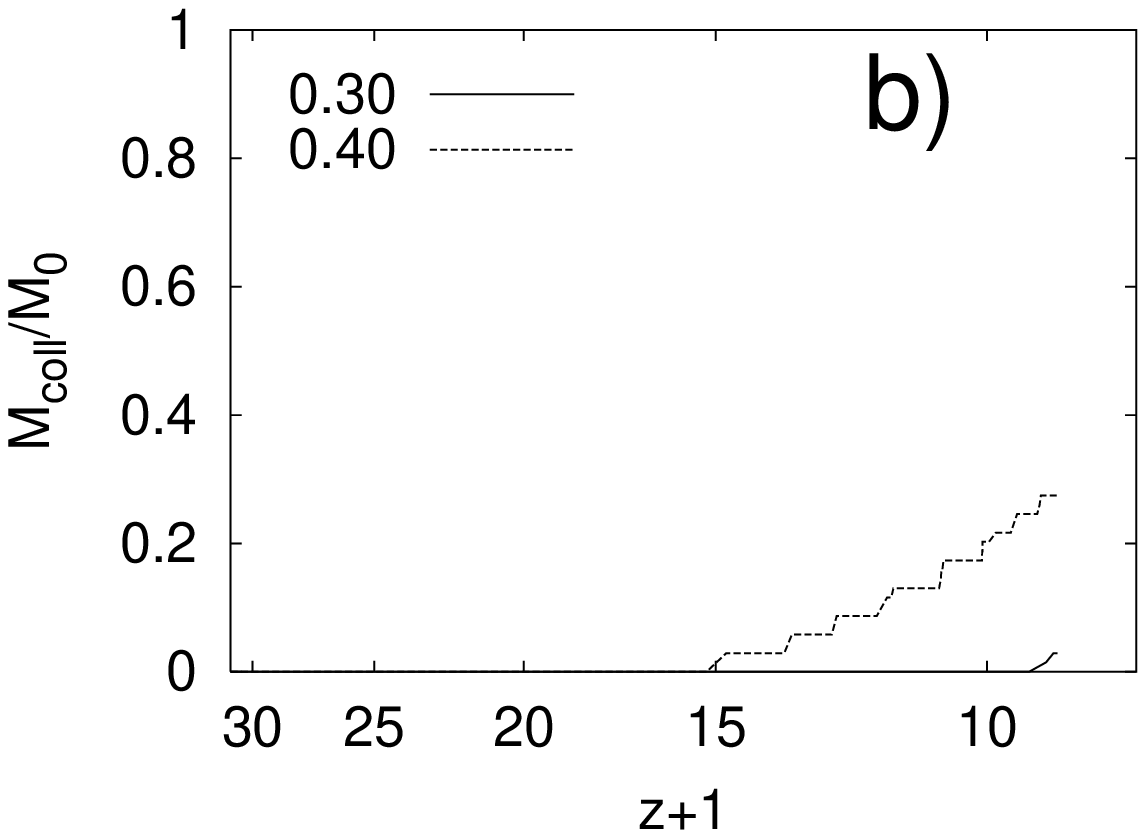}{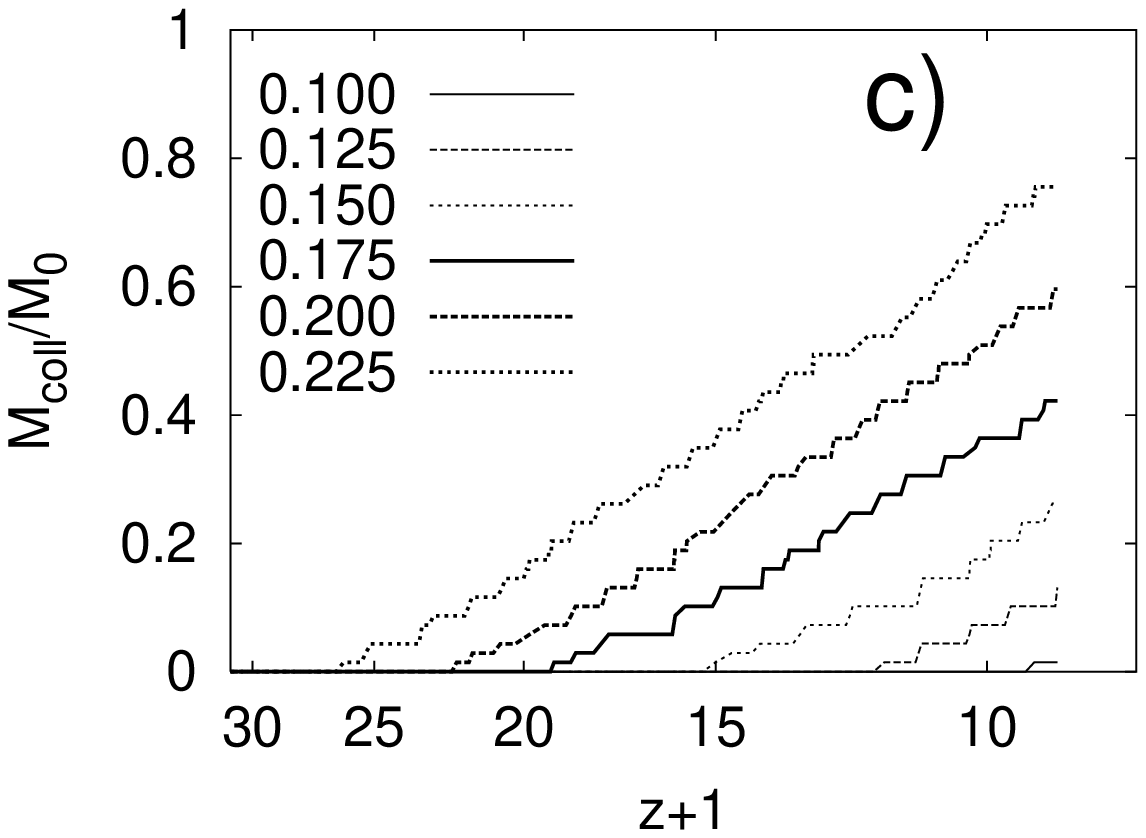}{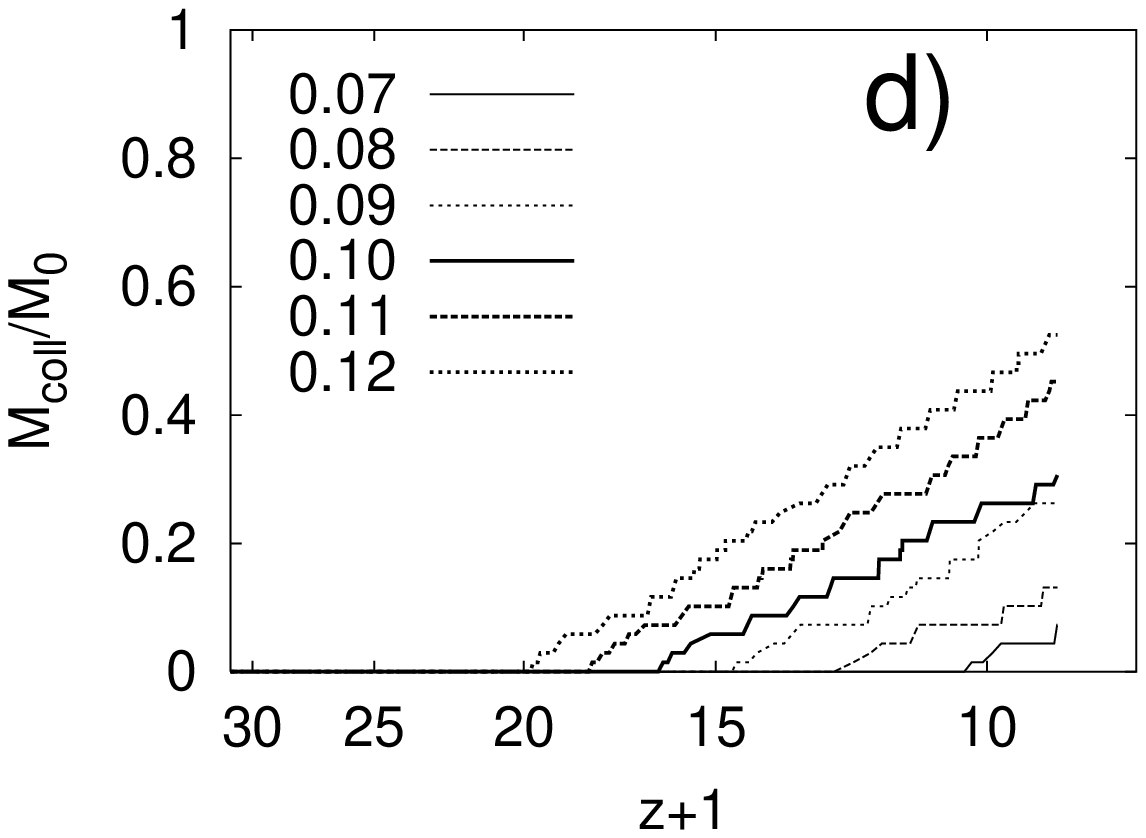}
{The collapsed mass fraction as a function of redshift for $M_0=2\times 10^3
M_{\odot}$, $3\times 10^3 M_{\odot}$, $2\times 10^4 M_{\odot}$ and
$1\times 10^5 M_{\odot}$ in the $\Lambda$-dominated Universe. Indicated
are overdinsities corresponding to different curves.}
{lambdacdm}

\poczwrys{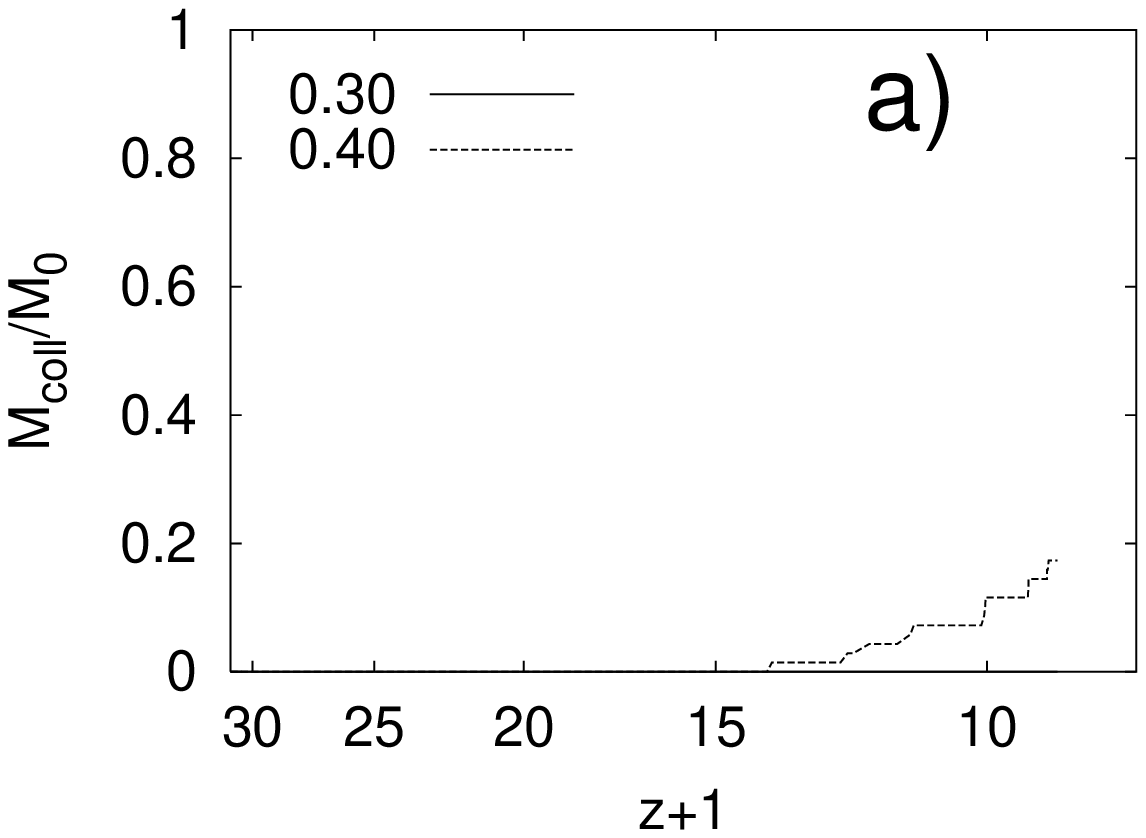}{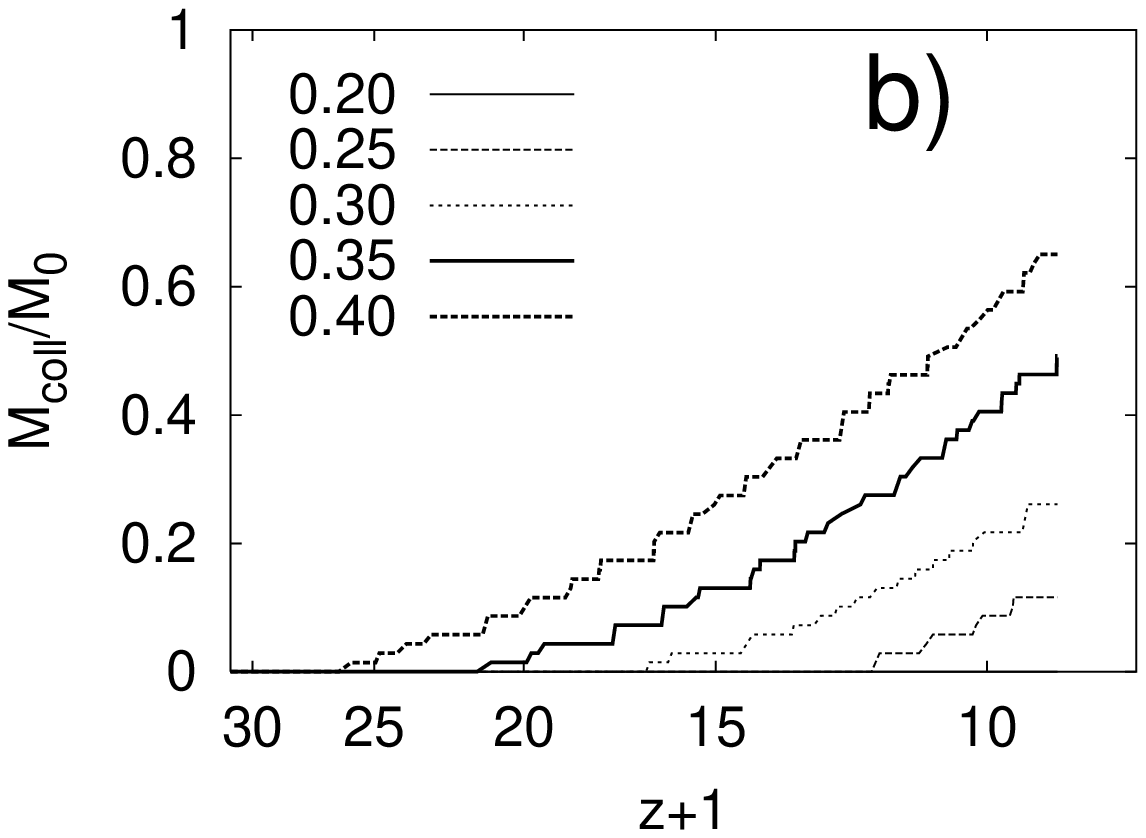}{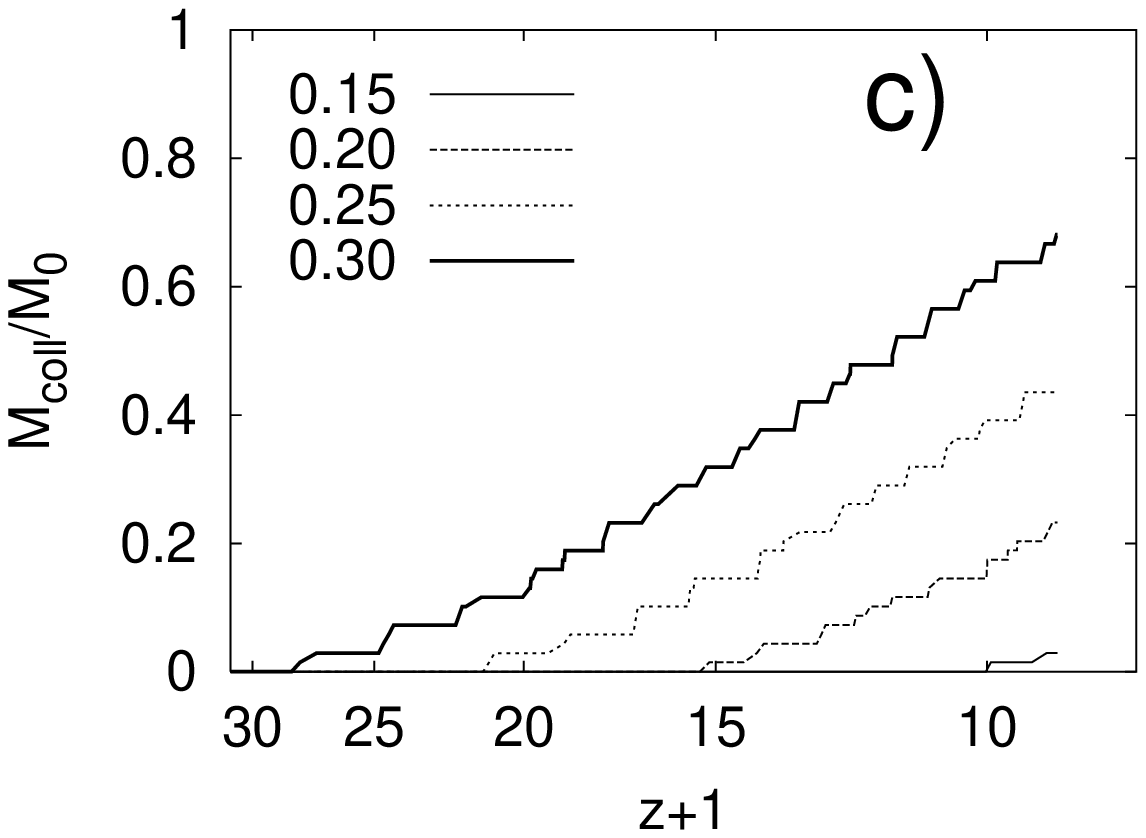}{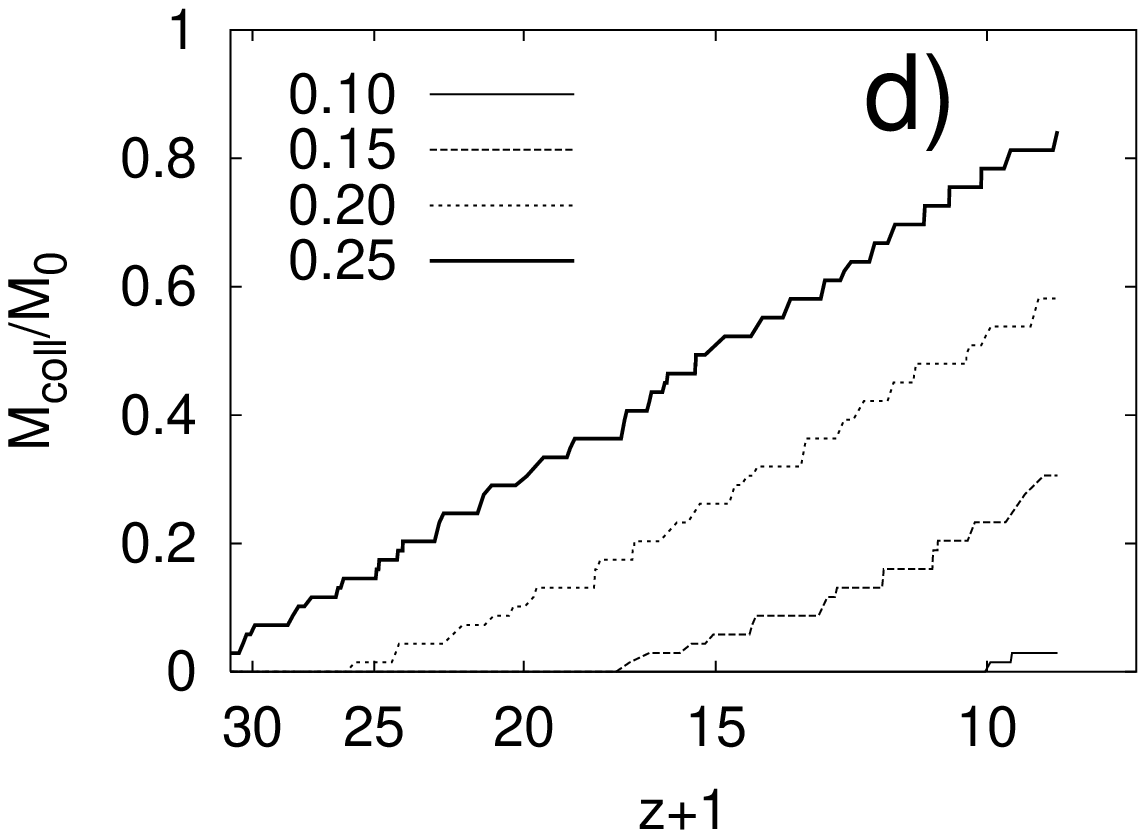}
{The same as in Fig. \ref{lambdacdm} for $M_0=5\times 10^2 M_{\odot}$,
$1\times 10^3 M_{\odot}$, $2\times 10^3 M_{\odot}$ and $5\times 10^3 M_{\odot}$
in the CDM-dominated Universe.}{scdm}

If there is a line that is described but cannot be found on a plot
it means that for this overdensity no baryonic shell has collapsed before
$z=8$.

The results confirmed our expectations. In the dark energy dominated Universe
initial clouds need greater mass and larger initial overdensities in order
to at least partially collapse before $z=8$. If we restrict ourselves to clouds
of $z=500$ DM overdensities not greater than 0.4, the minimal mass needed for
at least partial collapse before $z=8$ is 3000~M$_{\odot}$ (Fig.
\ref{lambdacdm} b). For $M=2000 M_{\odot}$ and $\delta_{dm,i}=0.4$ the fraction
of the collapsed mass is on order two percent (Fig. \ref{lambdacdm} a).
One should
remember that for these cloud masses the rms fluctuations at $z=500$ are about
0.05 so overdensities equal to 0.4 represent $8\sigma$ peaks, which are {\bf
extremely unlikely}.
For the dark matter dominated Universe, the minimal cloud mass is about 500
M$_{\odot}$ (Fig. \ref{scdm} a) -- for the initial overdensity equal to 0.4
(the rms overdensity is about 0.14) about 9\% of the cloud mass may collapse
before $z=8$.

If we look at higher masses, the collapse becomes easier and easier,
threshold initial overdensities for partial collapse before $z=8$ fall much
faster than the rms overdensities.
However, for the $\Lambda$-dominated Universe and cloud masses not greater
than 10$^4$ M$_{\odot}$ the clouds need overdensities
larger than $3 \sigma$ in order to at least partially collapse --
unlike in the flat CDM model. For greater baryonic masses the collapse is
a bit more likely but even for 10$^5$ M$_{\odot}$ (Fig. \ref{lambdacdm} d)
they need at least 2 $\sigma$ to start their collapse before $z=8$.

Fig. \ref{3sigma} shows approximately the  redshift when the most innermost
shell has collapsed for various masses and $3\sigma$ overdensities in the
$\Lambda$CDM Universe and for $2\sigma$ overdensities in the flat CDM
Universe.

\rysunek{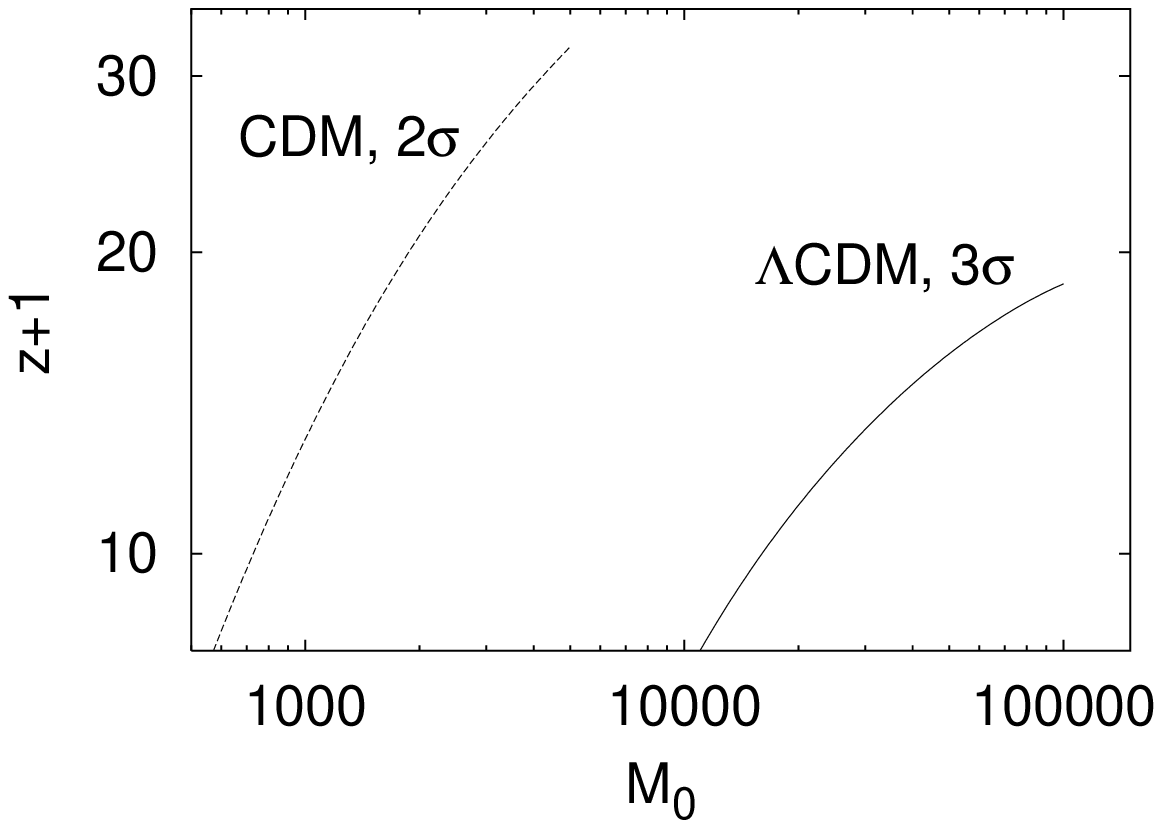}
{Approximate redshift when the object starts its collapse.}{3sigma}

In Figs. \ref{rew} a-d we compare the evolution of dark matter
shells and baryon
matter shells of the cloud with mass $M_0=5000 M_{\odot}$, for the
dark energy and dark matter dominated Universes. The values of initial
dark matter density enhancements are, respectively, $\delta_{dm,i}=0.30$
and $\delta_{dm,i}=0.20$.

\poczwrys{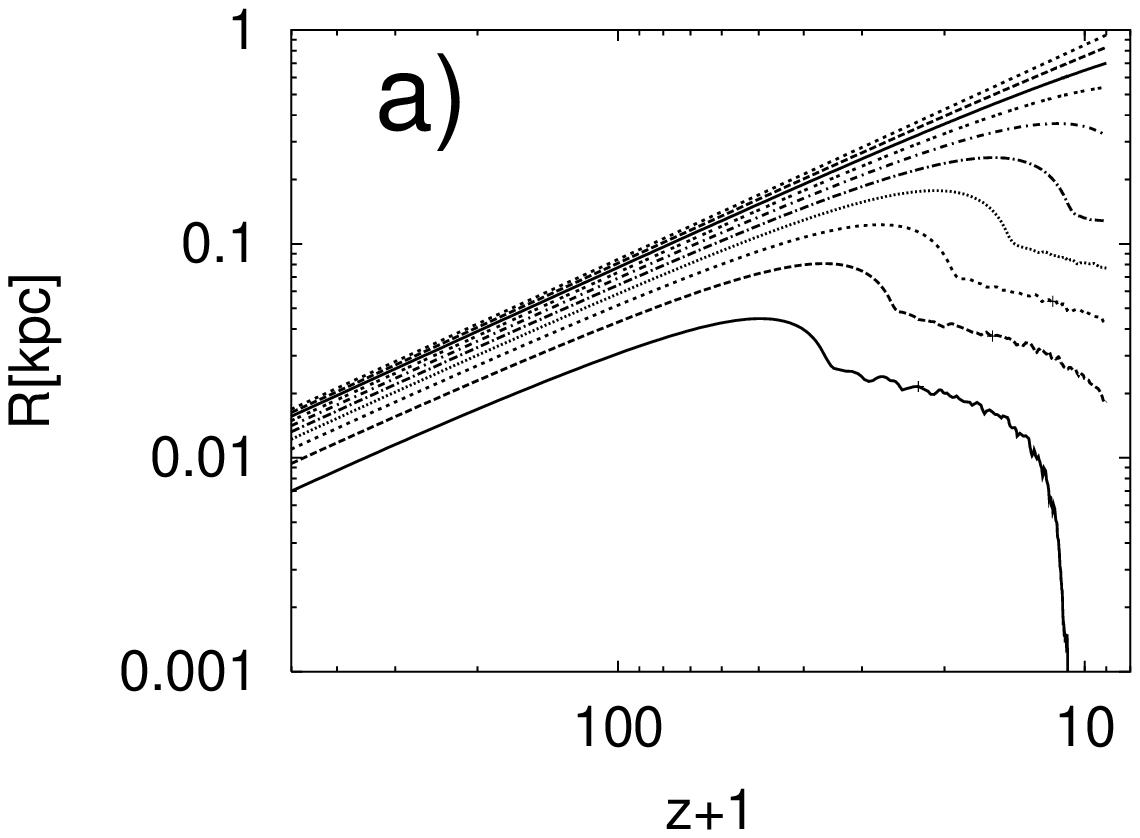}{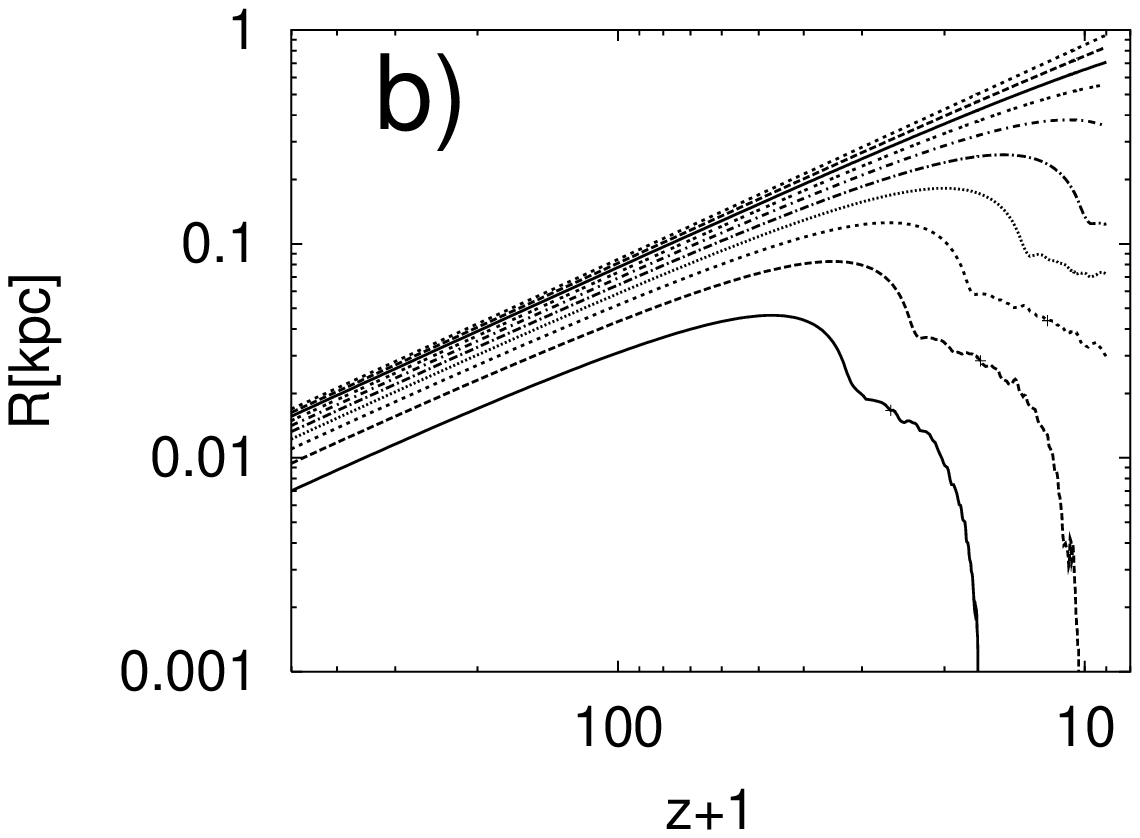}{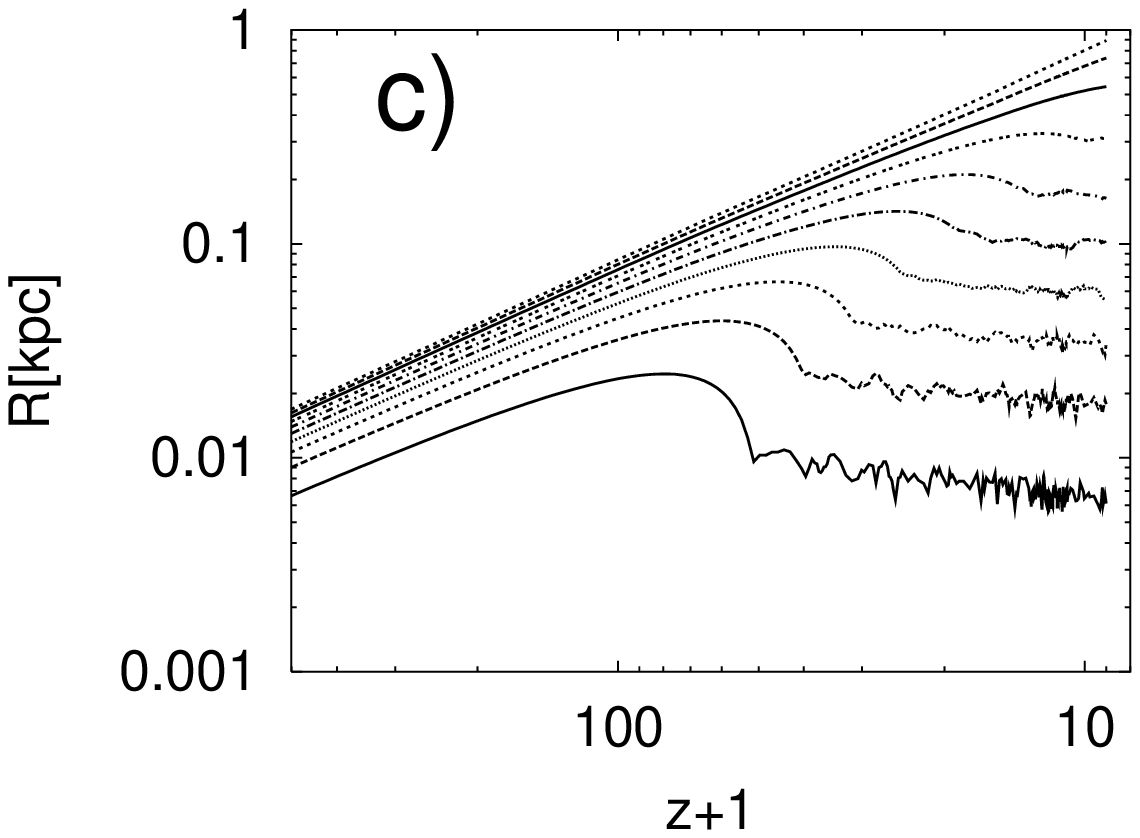}{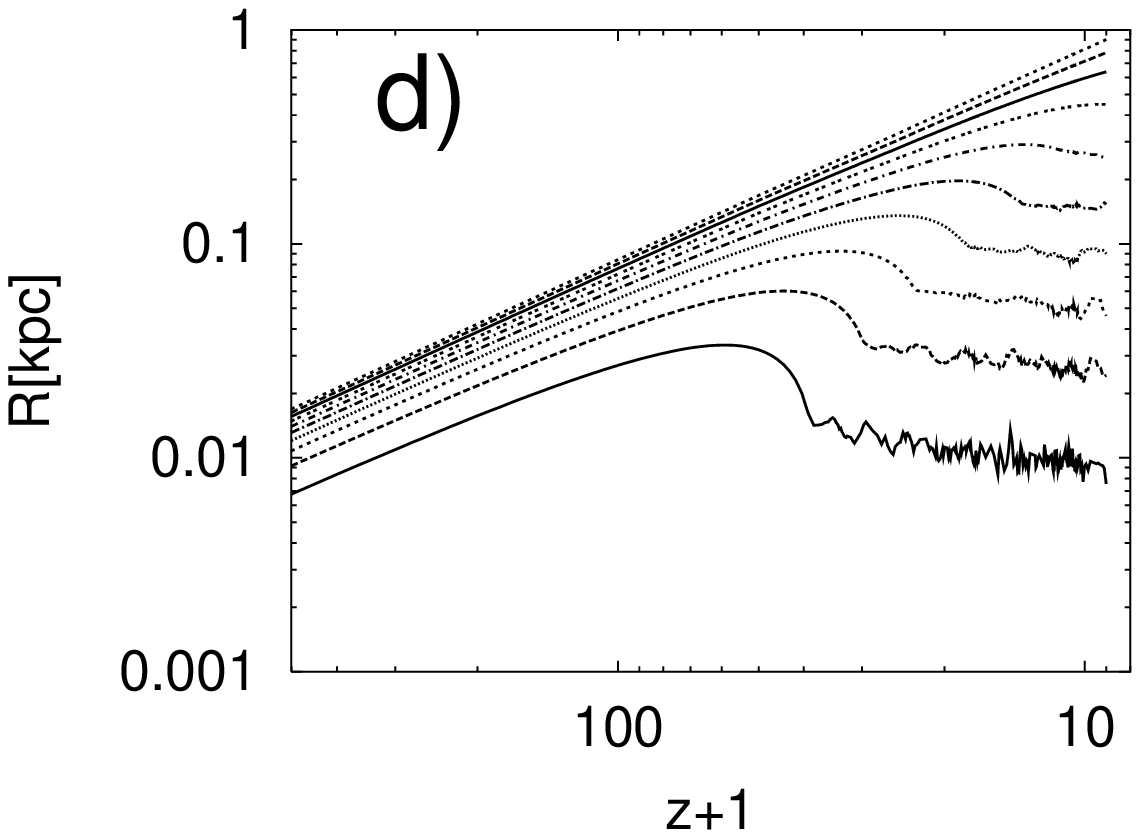}
{Evolution of the baryonic (upper plots) and dark matter (bottom plots)
shells for $M_0=5\times 10^3 M_{\odot}$, $\delta_{dm,i}=0.30$ for the
$\Lambda$-dominated Universe (left) and  for $M_0=5\times 10^3 M_{\odot}$,
$\delta_{dm,i}=0.20$ for the CDM-dominated Universe (right).}{rew}

The shells shown on the plots enclose $7\%$, $17\%,
27\%,...,97\%$ of the bound mass. This fractional division of mass applies
to both dark matter and baryon matter shells. It applies also to the
Fig. \ref{txew} (plots a and b) that show thermal evolution of gas shells.

The behaviour of both baryonic and dark components is quite typical, very
similar to the results of \citet{Hai96} and
our previous work \citep{Sta01}. Dark matter shells expand to some maximum
radius, then recollapse to about a half of this value and then very slowly
contract. In our calculations the behaviour of dark matter is quite
independent
of the cloud mass -- however, the position and the value of the maximal radius
depend on the initial overdensity. In contrast, the behaviour of baryon
matter shells strongly depends on the cloud
mass and initial overdensity. Of course, higher masses and overdensities
cause faster and more violent final collapse.

\poczwrys{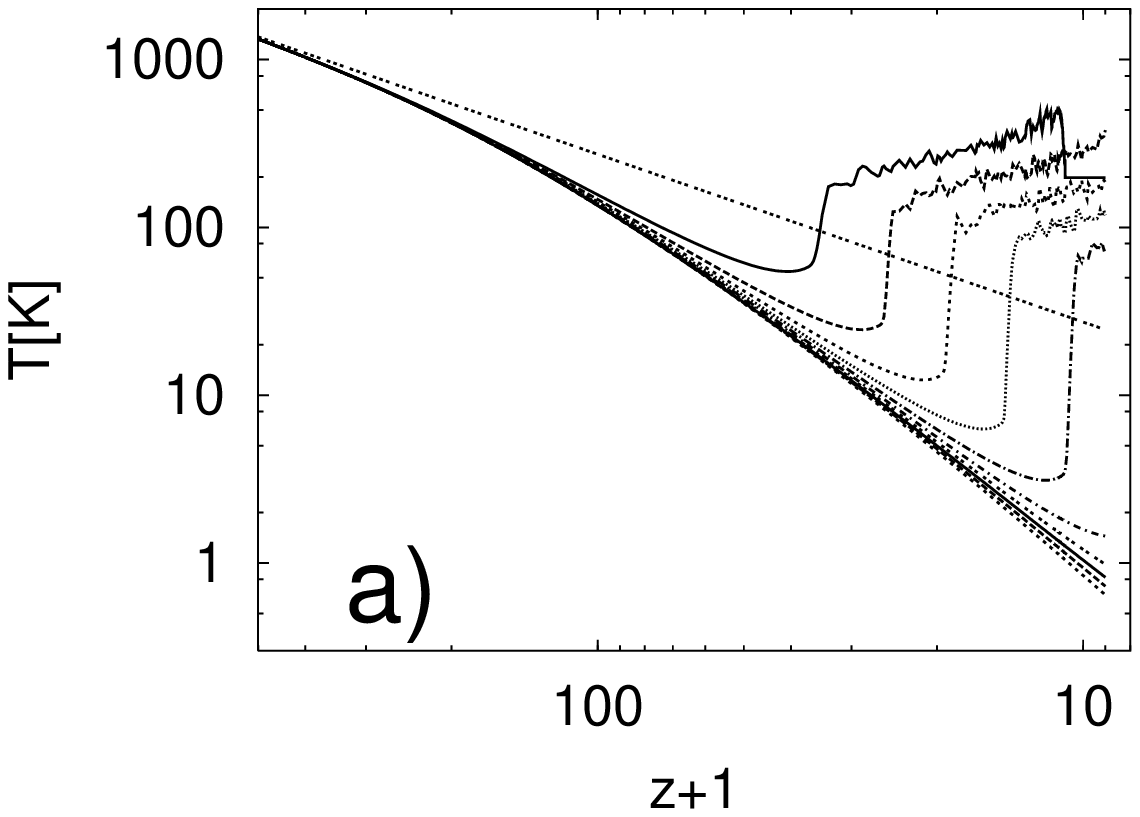}{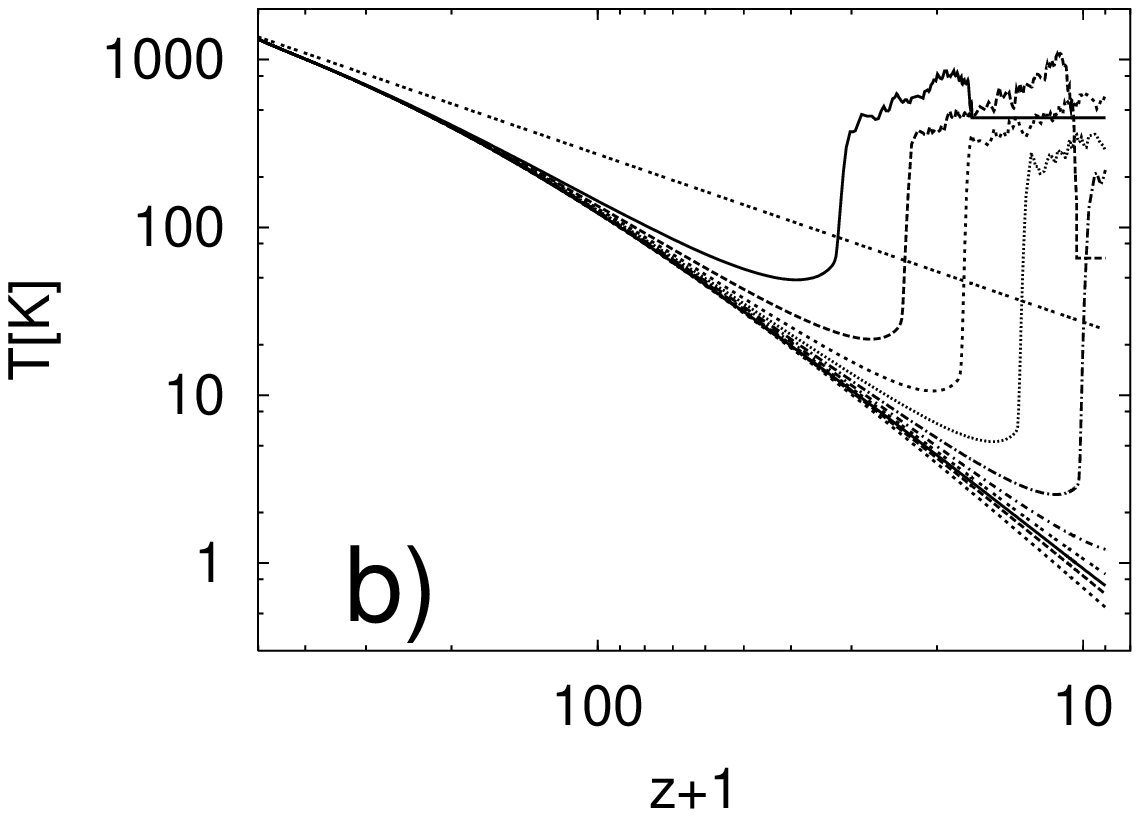}{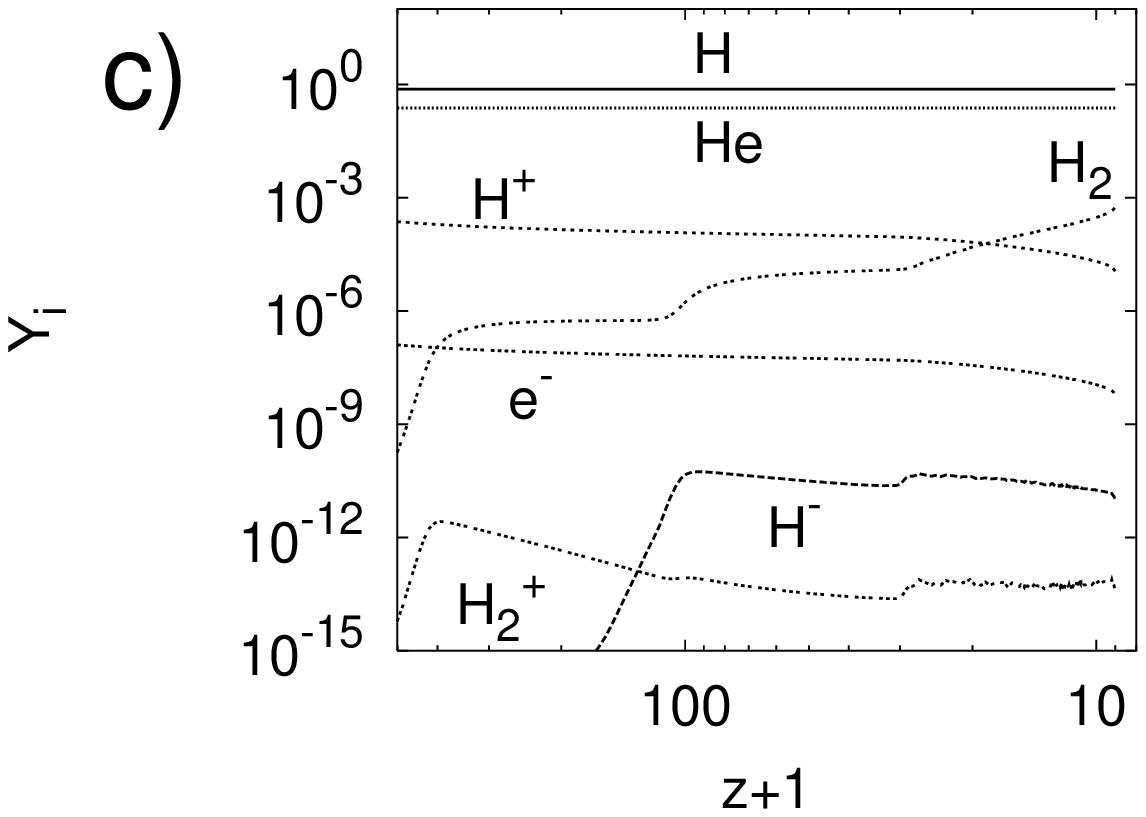}{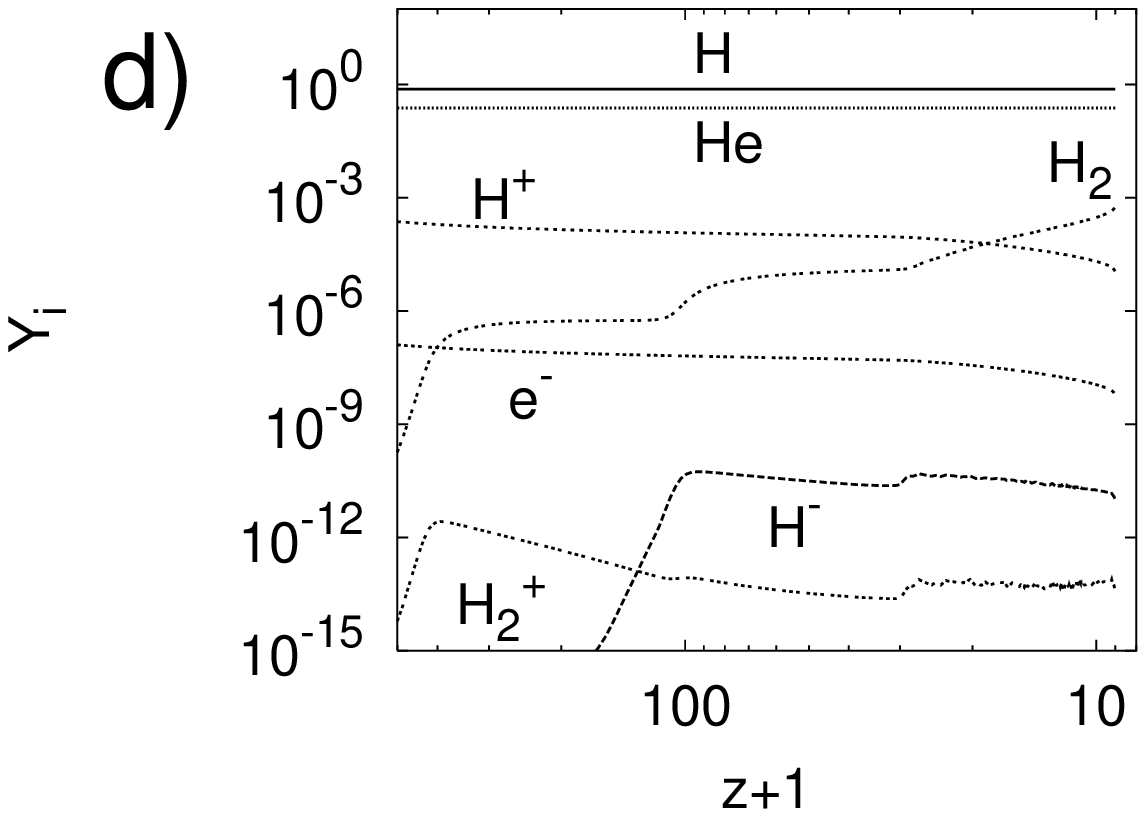}
{Evolution of the shell temperatures and chemical evolution for a shell
enclosing 12\% of the bound mass, for $M_0=5\times 10^3 M_{\odot}$,
$\delta_{dm,i}=0.30$ for the $\Lambda$-dominated Universe (left) and for
$M_0=5\times 10^3 M_{\odot}$, $\delta_{dm,i}=0.20$ for the CDM-dominated
Universe (right).}{txew}

In Figs. \ref{txew} a and b we show the temperature of shells as a
function of redshift. The curves indicating temperatures of baryonic shells
behave in an opposite way than the radii of baryonic
shells. The temperature  falls during expansion and  increases
in the collapse phase. After virialization the temperature
remains roughly constant until it starts to fall quite rapidly during the
final collapse. These figures show that virial temperatures in Standard CDM
are higher than in $\Lambda$CDM for objects of the same mass: even though for
the $\Lambda$CDM cloud initial overdensity was higher (0.3 compared with 0.2),
virial temperatures of its shells were about two times smaller. We also have
noticed that in both cosmological models shells that undego the final collapse
must reach some critical temperature.
For $\Lambda$CDM it is about 500 K while for Standard CDM it is about 450 K
and it is quite independent on the mass of the cloud. Shells at lower
temperatures do not collapse before $z=8$.

The abundance  (by mass) of various species as a function of redshift is shown
in Figs. \ref{txew} c and d. The results correspond to the shell of mass
$M=0.12 M_z$ of the same cloud as in Figs. \ref{rew} and \ref{txew} a, b. One
can notice the increase of the amount of H$_2$ molecules at later redshift
when the shell collapses. This makes H$_2$ cooling more
efficient, triggering the final collapse of inner shells.

\poszrys{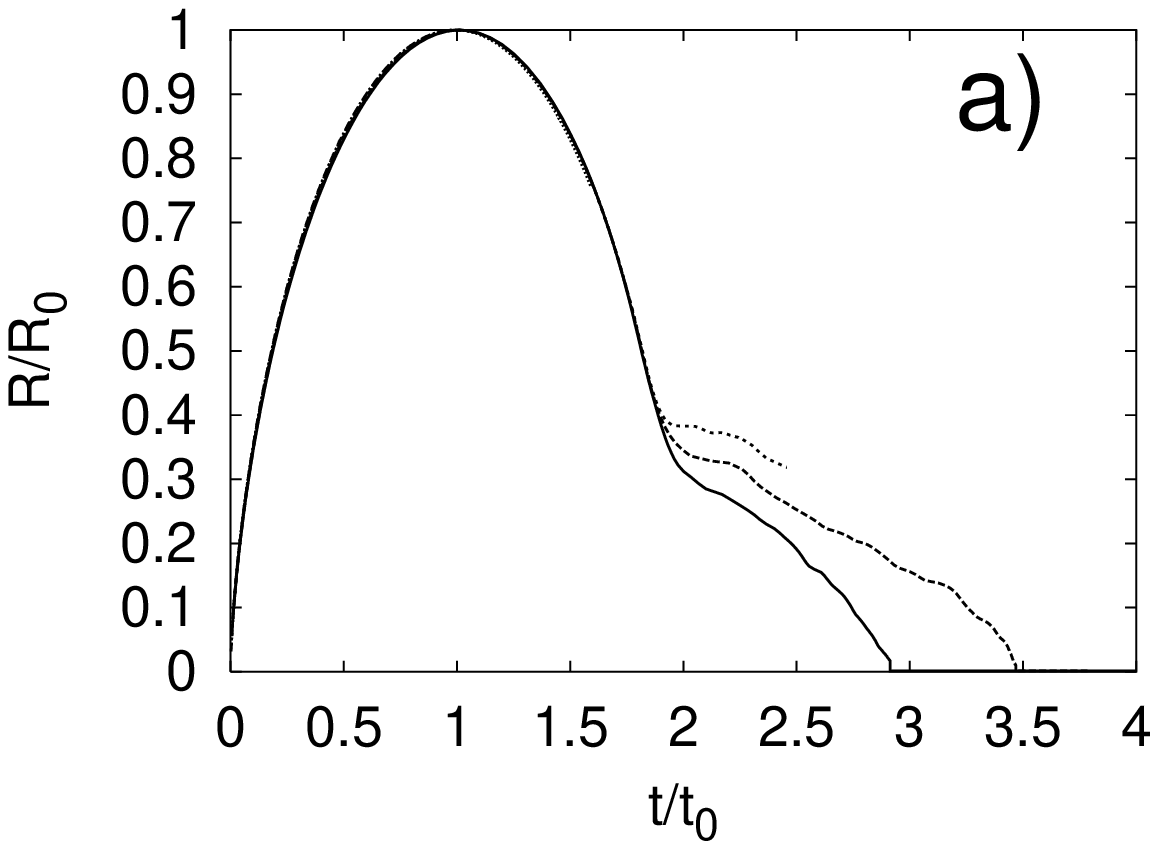}{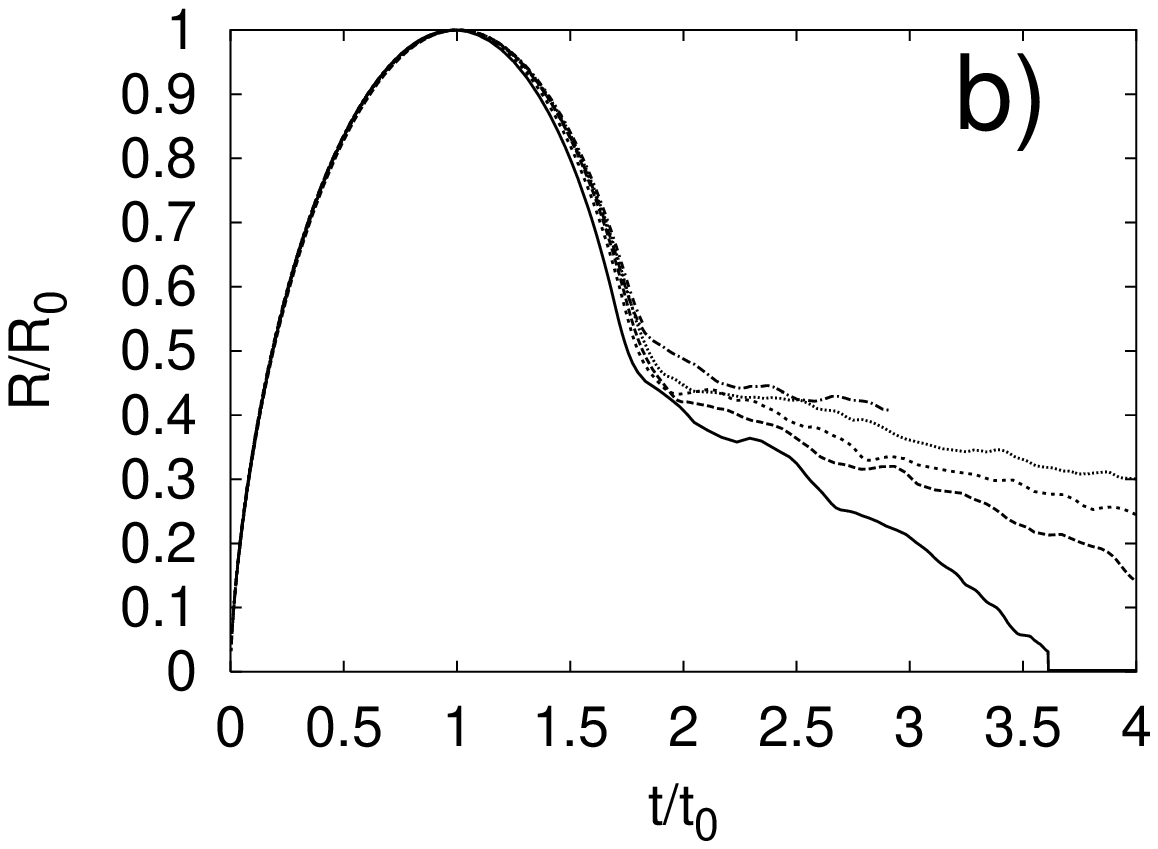}{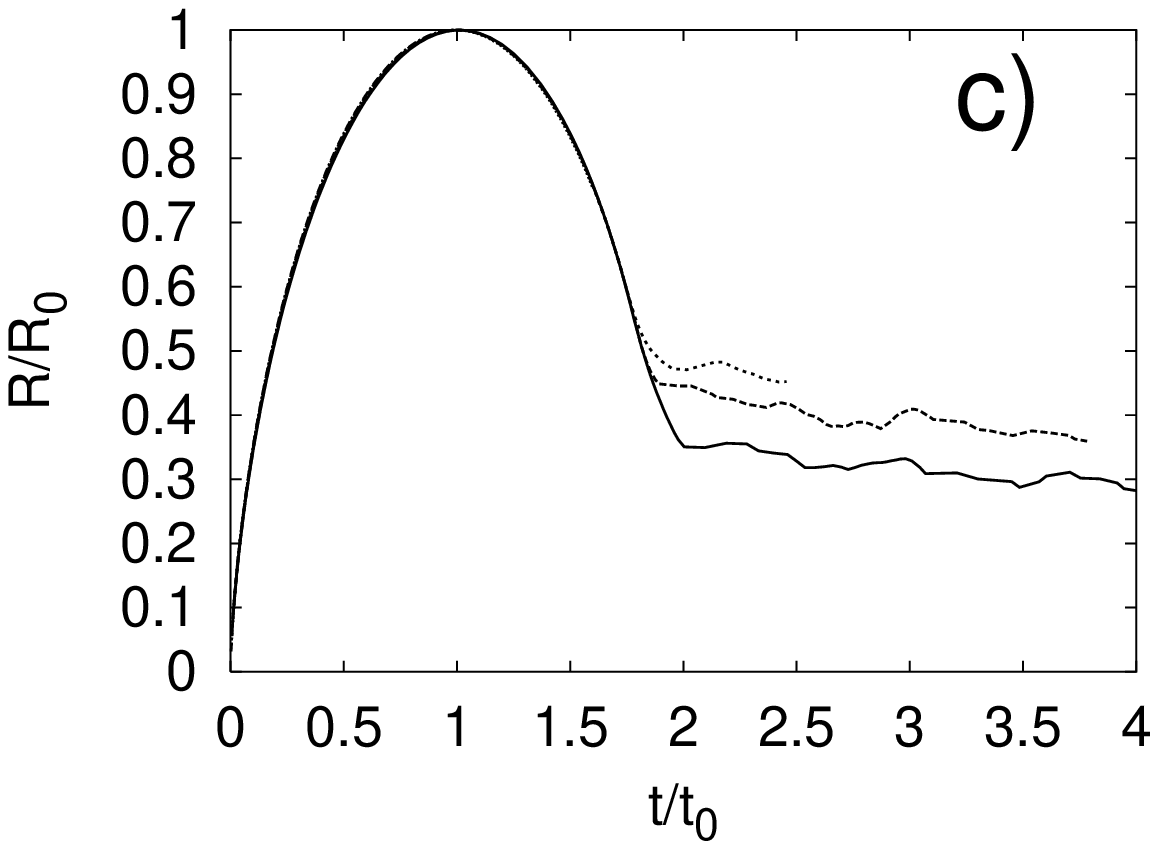}{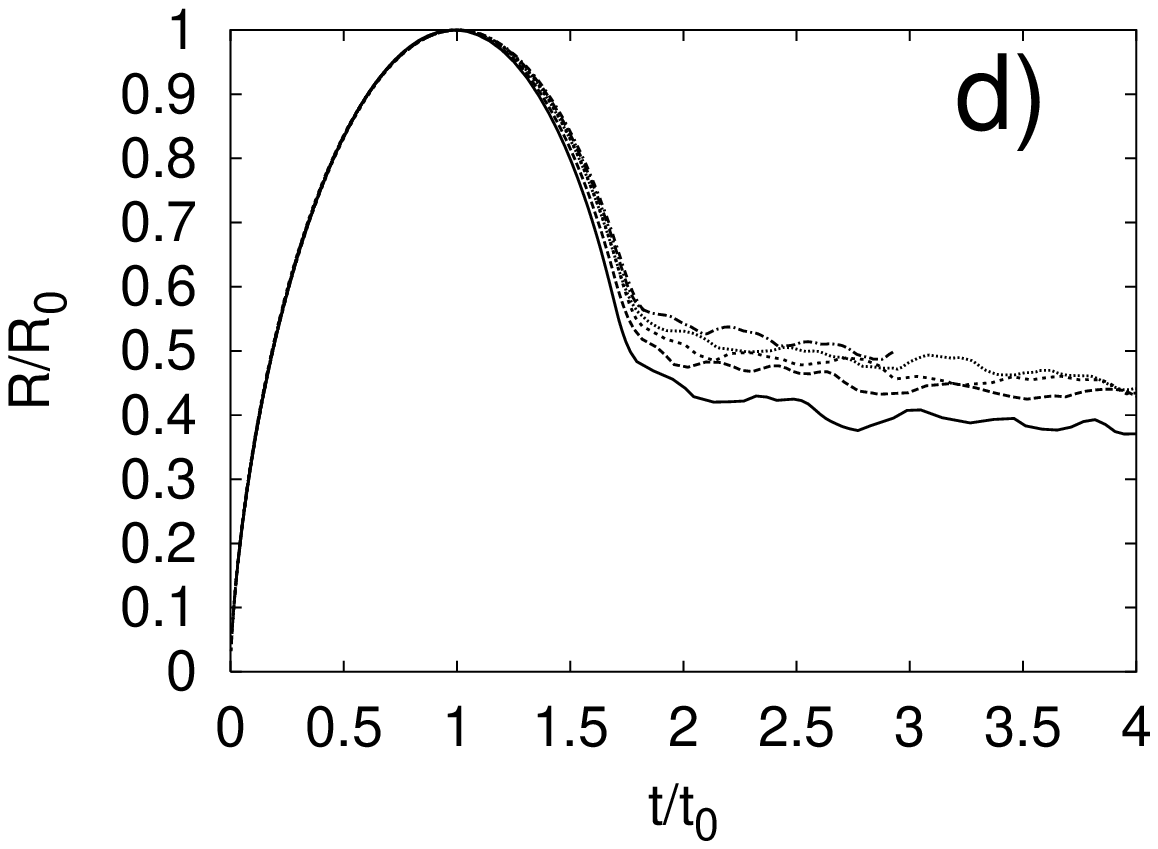}{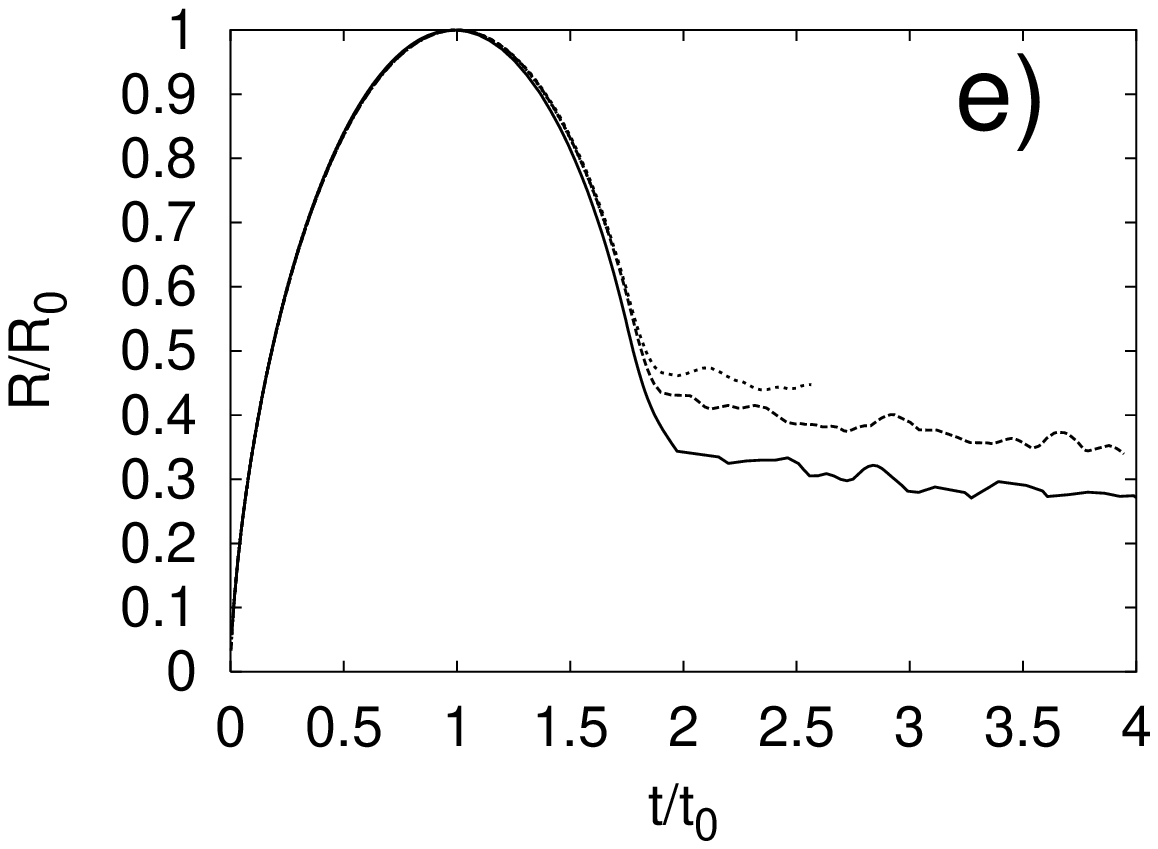}{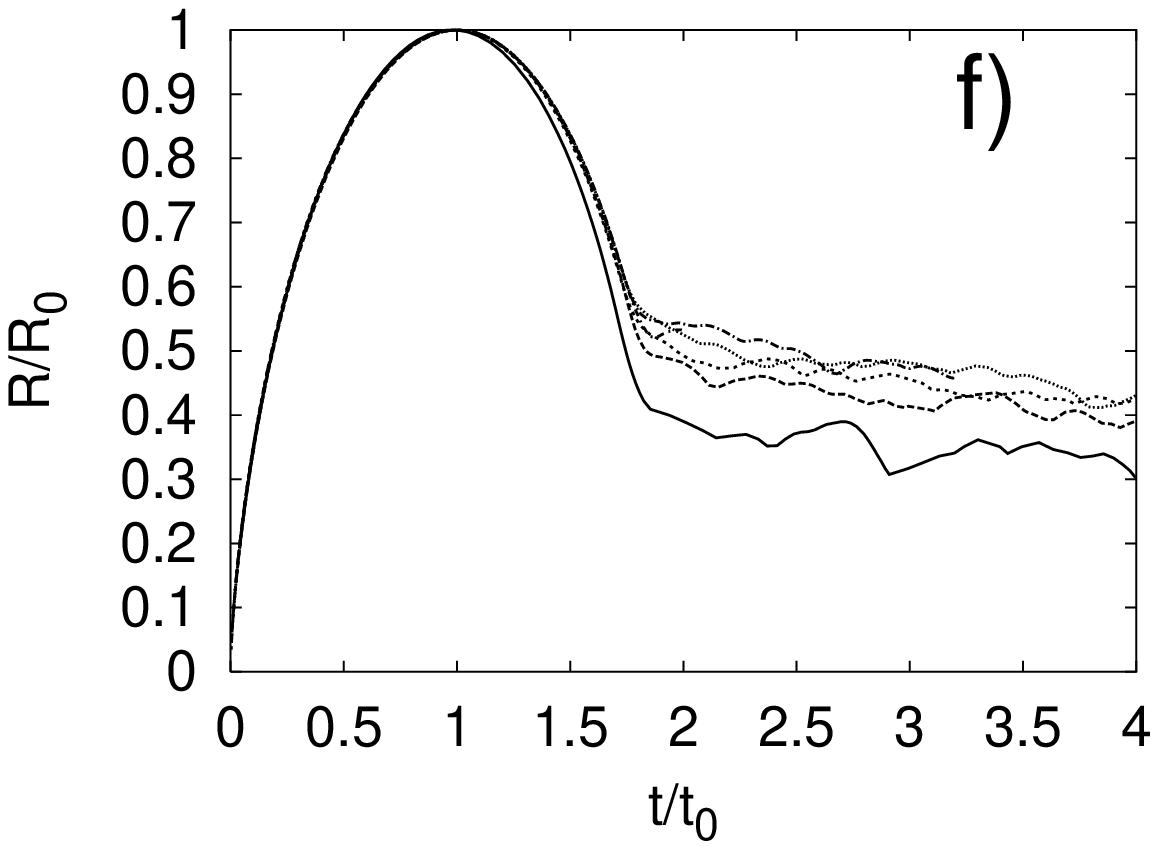}
{Normalized evolution of baryonic shells for $M_0=5\times
10^3 M_{\odot}$, $\delta_{dm,i}=0.30$ for the $\Lambda$-dominated
Universe (left) and for $M_0=5\times 10^3 M_{\odot}$, $\delta_{dm,i}=0.25$ for
the CDM-dominated Universe (right) with cooling (upper plots), without H$_2$
cooling (plots in the middle) and without any cooling or heating (bottom
plots).}
{selfsimilar}

In Figs. \ref{selfsimilar} a-f we show plots similar to Fig. \ref{rew} but
normalized to the maximum radius R$_0$ and time of the turnaround t$_0$
for $M_0=5\times 10^3 M_{\odot}$, $\delta_{dm,i}=0.30$ for the
$\Lambda$-dominated Universe (plots a, c and e) and  for
$M_0=5\times 10^3 M_{\odot}$, $\delta_{dm,i}=0.25$ for the CDM-dominated
Universe (plots b, d and f) with cooling, without H$_2$ cooling and without
any cooling or heating (including photoionization). While plots with no
cooling/heating (e and f) and no H$_2$ cooling (c and d) are almost
self-similar for different shells (the difference in final radii is probably
due to different mean initial overdensity for each shell) and similar to those
predicted by Bertschinger \citep{Ber85}, plots with cooling are self-similar
and Bertschinger-like up to virialization only. It also shows that H$_2$
cooling is necessary to collapse before $z=8$.

\section{Conclusions}

For the dark energy dominated Universe we need very large overdensities,
usually much exceeding the 3$\sigma$ limit, for low-mass
($M_0 \leq 10^5 M_{\odot}$) objects in order to at least partially collapse
before $z=8$. For the flat, dark matter dominated Universe the collapse is
much faster and direct formation of a low-mass object is quite likely,
even as soon as for $z\sim 20 - 30$. It is a result of much lower
normalization in $\Lambda$CDM (Table 1) and lower virial
temperatures for $\Lambda$CDM clouds of the same mass and initial overdensity.
A secondary reason may
be that in the flat CDM model any overdensity will stop its evolution and
possibly cool and collapse while for $\Omega_M < 1$ there is some threshold
overdensity that depends on the model and epoch.
The chemical evolution is pretty much
insensitive to cosmology, at least before the final collapse.
In the $\Lambda$-dominated Universe perhaps low-mass objects may form through
fragmentation of greater objects. Direct formation of such objects before
$z\sim 10$ seems  unlikely.

\end{document}